\documentclass[useAMS,usenatbib]{mn2e}

\usepackage{graphicx}
\usepackage{natbib}

\usepackage{txfonts}
\title[The NGC7448 region]{The Arecibo Galaxy Environments survey IV: the NGC7448 region and the HI mass function}
\author[Davies et al.]
{J. I. Davies$^{1}$, R. Auld$^{1}$,  L. Burns$^{1}$, R. Minchin$^{2}$, E. Momjian$^{3}$, S. Schneider$^{4}$,  \newauthor
M. Smith$^{1}$, R. Taylor$^{1}$, W. van Driel$^{5}$\\
$^{1}$School of Physics \& Astronomy, Cardiff University, Queens Buildings The Parade, Cardiff CF24 3AA, UK. \\
$^{2}$Arecibo Observatory, HC03 Box 53995, Arecibo, PR 00612, USA.    \\
$^{3}$NRAO, Dominici Science Operations Center, P.O. Box O, 1003 Lopezville Road, Socorro, NM 87801, USA.    \\
$^{4}$Department of Astronomy, University of Massachusetts, 710 North Pleasant Street, Amherst, MA 01003, USA.   \\
$^{5}$GEPI, Observatoire de Paris, CNRS, Universit\'e Paris Diderot, 5 place 
Jules Janssen, 92190 Meudon, France  \\ 
 }

\begin{document}

\date{Original October 2010}


\maketitle


\begin{abstract}
In this paper we describe results from the Arecibo Galaxy Environments Survey (AGES). The survey reaches column densities of $\sim3\times 10^{18}$ cm$^{-2}$ and masses of $\sim10^{7}$ M$_{\odot}$, over individual regions of order 10 sq deg in size, out to a maximum velocity of 18,000 km s$^{-1}$. Each surveyed region is centred on a nearby galaxy, group or cluster, in this instance the NGC7448 group. Galaxy interactions in the NGC7448 group reveal themselves through the identification of tidal tails and bridges. We find $\sim2.5$ times more atomic gas in the inter-galactic medium than in the group galaxies. We identify five new dwarf galaxies, two of which appear to be members of the NGC7448 group. This is too few, by roughly an order of magnitude, dwarf galaxies to reconcile observation with theoretical predictions of galaxy formation models. If they had observed this region of sky previous wide area blind HI surveys, HIPASS and ALFALFA, would have detected only 5\% and 43\% respectively of the galaxies we detect, missing a large fraction of the atomic gas in this volume. We combine the data from this paper with that from our other AGES papers (370 galaxies) to derive a HI mass function with the following Schechter function parameters $\alpha=-1.52 (\pm 0.05)$, $M^{*}=5.1 (\pm 0.3) \times 10^{9}$ h$_{72}^{-2}$ M$_{\odot}$, $\phi=8.6 (\pm 1.1) \times 10^{-3} $ h$^{3}_{72}$ Mpc$^{-3}$ dex$^{-1}$. Integrating the mass function leads to a cosmic mass density of atomic hydrogen of $\Omega_{HI}=5.3 (\pm 0.8) \times 10^{-4}$ h$^{-1}_{72}$. Our mass function is steeper than that found by both HIPASS and ALFALFA ($\alpha=1.37$ and 1.33 respectively), while our cosmic mass density is consistent with ALFALFA, but 1.7 times larger than found by HIPASS.
\end{abstract}


\section{Introduction} 
Equipped with its multi-beam instrument (Arecibo L-band Feed Array - ALFA) the 305 metre Arecibo telescope is the state of the art facility for carrying out extended blind HI surveys of large areas of sky. As part of the Arecibo Galaxy Environments Survey (AGES) we have been using ALFA to survey selected regions of the sky. These regions are centred on relatively nearby galaxies, galaxy groups or clusters (Paper I, Auld, et al. 2006, Paper II, Cortese et al. 2008, Paper III, Minchin et al. 2010, Taylor 2010). Our primary objective is to reach low HI column densities ($\approx 3\times10^{18}$ cm$^{-2}$) and masses ($\approx 10^{7}$ M$_{\odot}$) to discover tidal debris, isolated HI clouds, dwarf and HI rich galaxies. With this information we will be able to measure more accurately the mass density of atomic gas in the Universe. There are two parts to the survey: the detailed analysis of the target galaxy, group or cluster and the discovery and measurement of galaxies detected in the surveyed volume. In this paper the nearby object is the NGC7448 galaxy group and we survey an area of approximately $3\times5$ sq deg out to a velocity of about 18,000 km s$^{-1}$. 
The AGES is fully described in Auld et al. (2006) with updates on our web site at http://www.naic.edu/$\sim$ages.

It is clear that environment can have a dramatic effect on the gas content of galaxies. For example galaxy clusters like Virgo have many galaxies that are deficient in HI compared to similar galaxies in the field (Chamaraux et al., 1980, Haynes and Giovanelli, 1986, Solanes et al., 2001). The group environment may also have an important role to play in the way galaxies evolve and
what makes galaxy groups particularly interesting is that at the current epoch the majority of the galaxy population resides within them (Tully 1987).  Due to this expected environmental effect, groups of galaxies have been the subject of many HI observations in the past, for example see: Haynes (1981), van Driel et al. (1992), Yun, Ho \& Lo (1994), Zwaan (2001), de Blok et al. (2002), Pisano et al. (2007), Chynoweth et al. (2009). In an extensive study of 15 nearby galaxy groups, Haynes (1981) found 6 that showed clear signs of tidal interactions (tidal streams) demonstrating the influence of the group environment. In particular in the NGC7448 group van Driel et al. (1992) found extended HI gas between the main optically bright galaxies. In total this inter-galactic HI amounted to some $1.2\times10^{10}$ M$_{\odot}$ of gas, 1.5 times that found within the galaxies. Contrary to this, Zwaan (2001) concludes, after studying five nearby groups similar to the Local Group, that the total amount of inter-galactic atomic gas in groups is less than 10\% of that in the galaxies. Clearly the true state of affairs is that the amount of hydrogen drawn out of galaxies in the group environment is quite variable, depending on the relative positions, masses and velocities of the galaxies involved.

The NGC7448 group sits in the local void south of the Galactic plane ($l\approx87$ deg and $b \approx -39$ deg, $A_{B}=0.36$, Schlegel et al. 1998). The NASA Extra-galactic Database (NED) gives a mean redshift independent distance to NGC7448 of 28.6 Mpc (redshift of $v=2194$ km s$^{-1}$), which we will use in this paper as the group distance. The very loose grouping of galaxies that may be associated with the group, from their spatial association and similarity of velocities (2-3000 km s$^{-1}$), extends from NGC7479 in the south to UGC12350 in the north ($\approx4.3$ deg). Our observations cover the central region of the group which, consists of six primary galaxies NGC7448, NGC7463, NGC7464, NGC7465, UGC12313 and UGC12321 and about 1.7 degrees to the south two other bright galaxies, NGC7437 and UGC12308. These last two are probably group members as their radial velocities are very similar to that of NGC7448 ($\approx$2200 km s$^{-2}$), Fig. 1. NGC7463/4/5 form a compact sub-group and based on their disturbed nature van Driel et al. (1992) suggest that NGC7464 and NGC7465 are in the process of merging (see also Li and Seaquist, 1994). 

There is also HI emission extending from NGC7463/4/5 towards NGC7448 (Haynes, 1981). It seems clear that tidal interactions are playing a prominent role in the evolution of these galaxies. Understanding the consequences of the interaction though is complex. Atomic hydrogen is the fuel for star formation so its removal is bound to suppress continued star forming activity, but we also know that tidal interactions can induce high star formation rates (Kennicutt et al. 1987). The NGC7463/4/5 sub-system in particular shows signs of recent star formation activity (van Driel et al. 1992) and all three galaxies have been classified as having an ultra-violet excess (Takase 1980). 

In the following sections we will describe the data acquisition and reduction of AGES observations of the NGC7448 group. Our results include an analysis of HI within and outside the main group galaxies and the discovery of new dwarf galaxies. We consider galaxies detected in the volume in front of and behind NGC7448 and use these to derive a HI mass function and calculate a value for the cosmic HI mass density.
\begin{figure}
\centering
\includegraphics[scale=0.5]{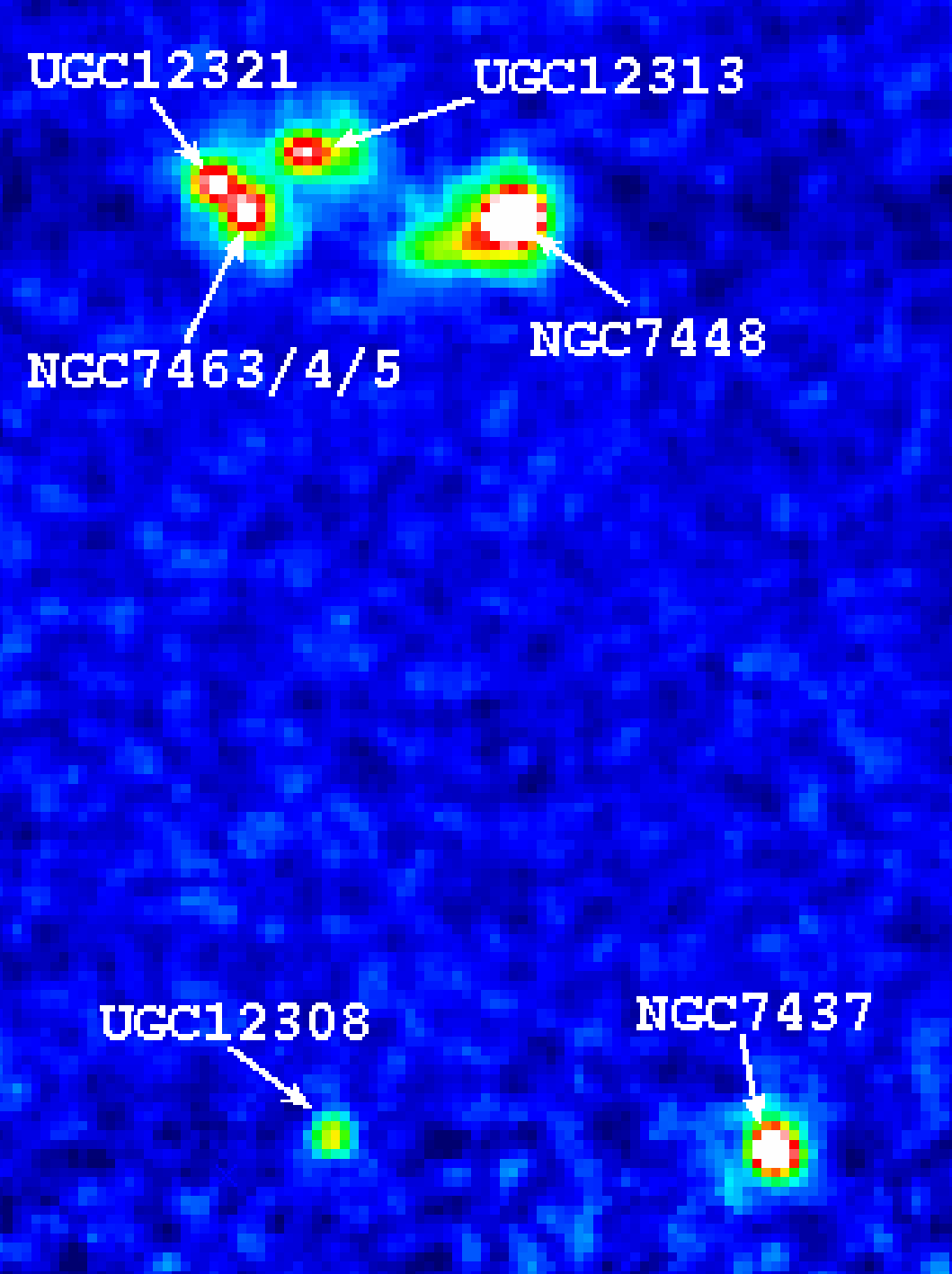}
\caption{This is an image of the maximum HI flux along each line of sight (pixel) between 1630 and 2550 km s$^{-1}$, to pick out the brightest sources irrespective of velocity. The image shows the brightest galaxies in the NGC7448 group. NGC7464 has the lowest velocity of 1875 km s$^{-1}$ and NGC7463 the highest of 2341 km s$^{-1}$. The three galaxies NGC7463/4/5 are too close together to be spatially resolved by the 3.5 arc min Arecibo beam. The image covers an area of $1.72\times2.37$ sq deg ($0.86\times1.19$ Mpc at the group distance) centred at RA(J2000)=23 00 21, Dec(J2000)=15 09 29, north at the top and east to the left.
}
\end{figure}

\section{Observations and data reduction}
The NGC7448 group was observed over a number of months in 2009. Approximately 120
hours was used to cover an area of $\sim 35$ sq degrees (1.5 x 2.5 Mpc) centred on RA=23
00 00, Dec=15 14 00 (J2000). We used the 7-feed ALFA multi-beam receiver and the
WAPP spectrometers to record two linear polarisations from each beam within a 100
MHz bandwidth. The angular resolution determined by the size of the ALFA beams is
$\approx 3.5$ arc min. We used the drift-scan observing mode, whereby the receiver is
held stationary and the sky drifts overhead. With this observing technique each
point on the sky takes about 12 sec to cross the beam (Giovanelli et al. 2005).
Twenty-five separate scans are then combined to give a total integration time of 300
seconds/beam. Total power is recorded every second for the seven beams, each polarisation
and the 4096 velocity channels of width 5.5 km/s. 
Data reduction is performed using LIVEDATA and GRIDZILLA (Barnes et al. 2001).
LIVEDATA carries out bandpass estimation and removal, Doppler tracking and flux
calibration. Various methods are available for estimating and removing the bandpass.
We found that the most appropriate settings for the AGES observations were to use
the 'Extended' method using the median estimator. The 'Extended' method is a
modification of the MINMED routine that was originally used for the Galactic high velocity cloud project
(Putman et al. 2002). With this modified routine, the user loads
the entire scan into memory and defines a sliding window which will define the
target for which the bandpass will be estimated. For each beam and polarisation the
entire scan outside of the target window is used to estimate the median value in
each channel in the spectrum. The 4096-channel spectrum, made up of these median
values, forms the estimate for the bandpass and is subtracted from the target
spectrum. 
The resultant baselines were found to be acceptably flat except in the case of very
strong continuum sources which produce large-amplitude standing waves between the
dish and the receiver. The waves introduce a quasi-sinusoidal ripple into the
baseline with a period characteristic of the light-travel time between the
receiver and the dish.

GRIDZILLA is used to combine the individual spectra to produce 3-D data cubes. The
gridding technique is exactly the same as for HI Parkes All Sky Survey (HIPASS) so the reader is referred to
Barnes et al. (2001) for full details. The default parameter values specific to the
Arecibo telescope were applied. The NGC7448 cube has 1 arc min pixels and the 4096
velocity channels have been Hanning smoothed to 10 km/s resolution. Making a
Nyquist-sampled map with every beam and then median combining results in a circular
beam with symmetrical sidelobes peaking at the 5-10\% level (Minchin et al. 2006). The
noise distribution is approximately Gaussian with a median rms value of 0.7 mJy when
measured along individual pixels.

Radio frequency interference introduces a constant source of man-made noise that can
far exceed the system noise over certain regions of the spectrum. For this data
there is constant interference at 1350 MHz (14620 km s$^{-1}$) and 1387.8 MHz (km s$^{-1}$)
and an intermittent source at 1310 MHz (8300 km s$^{-1}$). None of these affect our
observations of the NGC7448 group though they must be considered when extracting
sources from the extended volume behind NGC7448. The velocity range $-80 < V < 40$
km s$^{-1}$ was also excluded from the extended search volume due to the presence of
contaminating signal from the Milky Way. 

We have checked the data calibration by comparing our HI fluxes with those given in
the NASA Extra-galactic Database (NED), using an average where there is more than one value.
Excluding the six galaxies in the central part of the NGC7448 group, where the HI
from different galaxies is difficult to separate, there are nineteen
galaxies in our area listed in NED with HI fluxes. For these nineteen we have fitted
the spectra as in Cortese et al. (2008) using the MBSPECT routine from the MIRIAD
spectral line analysis package. With the exception of NGC7464 and NGC7465 (see below) all subsequent 
HI parameters described in this paper have been obtained in this same way. Given the numerous different sources for the
NED fluxes the comparison is reasonably good (Fig. 2), with a linear least squares
fit giving $F_{NED}=0.85\pm0.03F_{AGES}+0.36\pm0.20$ Jy km s$^{-1}$.
\begin{figure}
\centering
\includegraphics[scale=0.52]{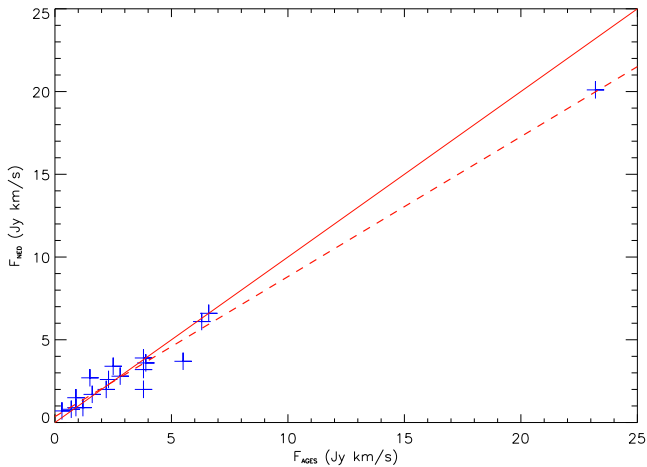}
\caption{A comparison of AGES HI line fluxes with those given in NED. The blue crosses indicate the data and have sizes approximately the size of the errors on each point (typically a 2 Jy km$^{-1}$ source has a flux error of about 0.3 Jy km$^{-1}$). The solid red line is the one to one relationship. The red dashed line is the linear least squares best fit to the data, F$_{NED}$=0.85F$_{AGES}$$\pm0.03$+0.36$\pm0.20$ Jy km s$^{-1}$.
}
\end{figure}

\section{The NGC7448 group}
The local environment can have a dramatic effect on a galaxy. For example in galaxy clusters gas can be stripped from a galaxy as it passes through the inter-galactic medium - ram pressure stripping (Gunn and Gott 1972, Vollmer et al. 2001). Galaxies can also be tidally buffeted as they move at relatively high speeds (600-700 km s$^{-1}$) through the varying cluster and galaxy gravitational potentials - harassment (Moore et al. 1996). Significant strong gravitational (tidal) interactions between individual galaxies are rare in clusters because typically they are moving past each other too quickly. This is not so for galaxy groups where the group velocity dispersion is of order the internal rotational velocity of the constituent galaxies. In this section we concentrate on the six bright galaxies in the central region of the NGC7448 group (Table 1). We will show that these galaxies have clear signs of gravitational interactions that have produced tidal streams and tails (see also Haynes 1981, van Driel et al. 1992) and that a large fraction of the total atomic gas now resides in the inter-galactic medium.

\begin{table*}
\begin{center}
\begin{tabular}{lcccccccc}
Name   &  RA  &  Dec(J2000)  & $V_{HI}$ (km s$^{-1}$) &   W$_{50}$ (km s$^{-1}$) &  W$_{20}$  (km s$^{-1}$) & $F_{Tot}$ (Jy km s$^{-1}$) &  $\frac{M_{HI}}{1.0\times10^{9}}$ M$_{\odot}$ &  $\frac{M_{Star}}{1.0\times10^{9}}$ M$_{\odot}$  \\ \hline
NGC7448 & 23 00 04 & 15 58 49 & 2192 & 271(252) & 297(365) & 24.2(23.3) & 4.8(4.8) & 24.0 \\
NGC7463 & 23 01 52 & 15 58 55 & 2343 & 189(227) & 220(246)  & 8.2(7.6)  & 1.6(1.6) & 22.4 \\
NGC7464 & 23 01 54 & 15 58 26 & 1943 & 177(234) & 549(245)  & 12.8(10.1) & 2.5(2.1) & 4.6 \\
NGC7465 & 23 02 01 & 15 57 53 & 1969 & 143(81) & 223(142)  & 15.0(11.7) & 2.9(2.3) & 10.6 \\
UGC12313 & 23 01 44 & 16  04 04 & 2018 & 123(139) & 210(177) & 6.11(9.8) & 1.2(2.0) & 0.7 \\
UGC12321 & 23 02 19 & 16 01 40 & 2162 & 163(205) & 188(231)  & 4.1(7.6) & 0.8(1.5) & 3.9 \\
DS96 J2259+1557      & 23 02 18 & 16 14 28 & 1994 & 89(82)       & 109       & 1.2(3.1)      & 0.2(0.5) & - \\
\end{tabular}
\caption{Properties of the central galaxies in the NGC7448 group. Numbers in brackets are those obtained by van Driel et al. (1992) except for DS96 J2259+1557 where they are the values given in Duprie and Schneider (1996). Stellar masses are calculated from the B band magnitudes given in Li and Seaquist (1994).}
\end{center}
\end{table*}

In this section we describe observations of the group galaxies and defer a discussion of sources detected throughout the data cube until section 5. The six group galaxies listed in Table 1 have a mean velocity of 2104 km s$^{-1}$ with a rms dispersion of 155 km s$^{-1}$. Their mean velocity width at 20\% peak flux density is 281 km s$^{-1}$. Thus the velocity dispersion of the group is of the same order as the rotational velocities of the galaxies and as such we might expect significant gravitational interaction (Toomre and Toomre, 1972, Haynes et al. 1984). In Fig 3 (top) we show a $1.6\times0.8$ sq deg region ($0.8\times0.4$ Mpc at the group distance) extracted from the HI data cube. The image has been made by taking the maximum pixel value along each line of velocity from 1641 to 2556 km s$^{-1}$. The image clearly picks out atomic gas that has been drawn out of the galaxies during the course of their gravitational interaction. There is a clear tail that has been drawn out of NGC7448 and a stream apparently extending from the NGC7463/4/5 sub-group towards NGC7448. In Fig. 3 (bottom) we show how this gas extends both spatially and in velocity.  There also appears to be gas connecting both spatially and in velocity NGC7463/4/5 and UGC12313, this 'bridge' of gas was originally detected in the higher resolution VLA observations of Li and Seaquist (1994), who suggest that it is due to a recent ($\le 10^{8}$ yr ago) close encounter between UGC12313 and NGC7463/4/5.

\begin{figure}
\centering
\includegraphics[scale=0.65,angle=-0]{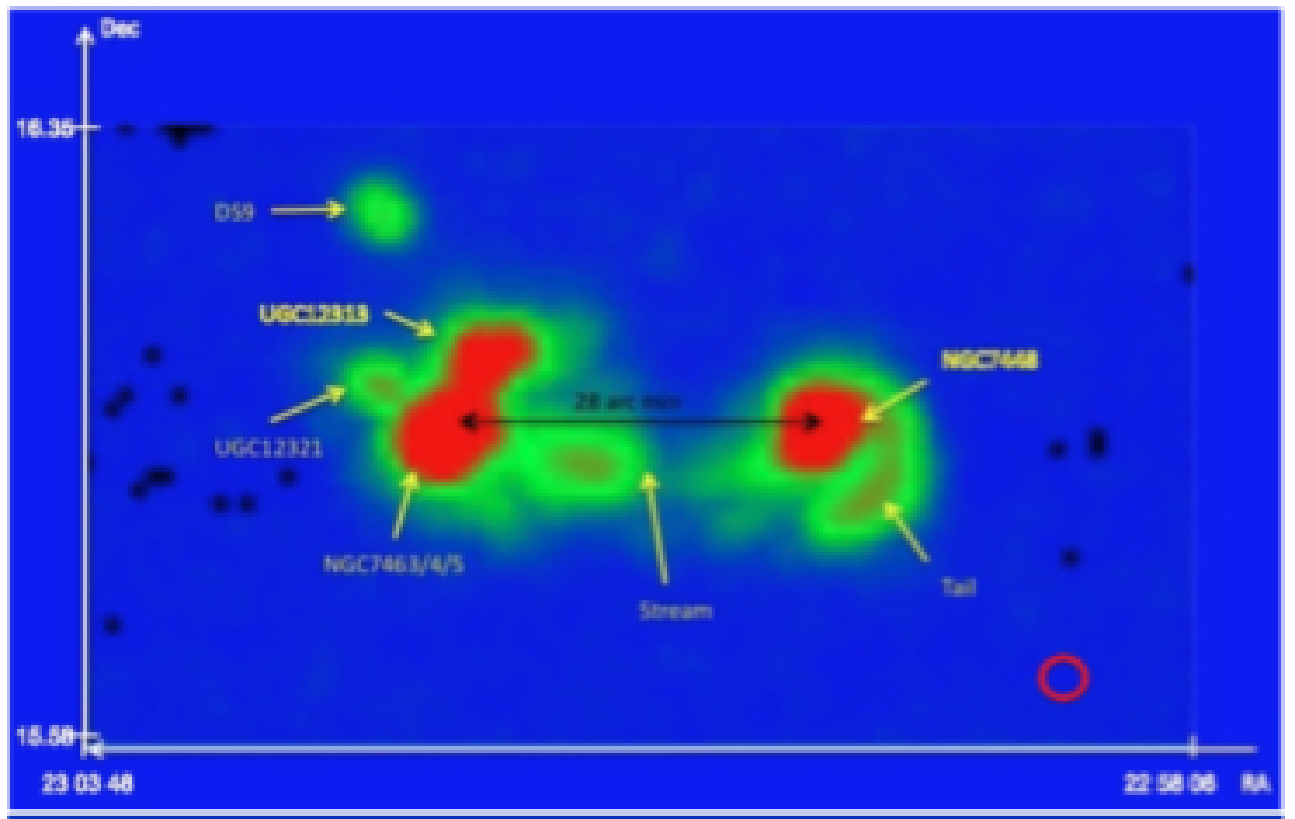}
\includegraphics[scale=0.258,angle=-0]{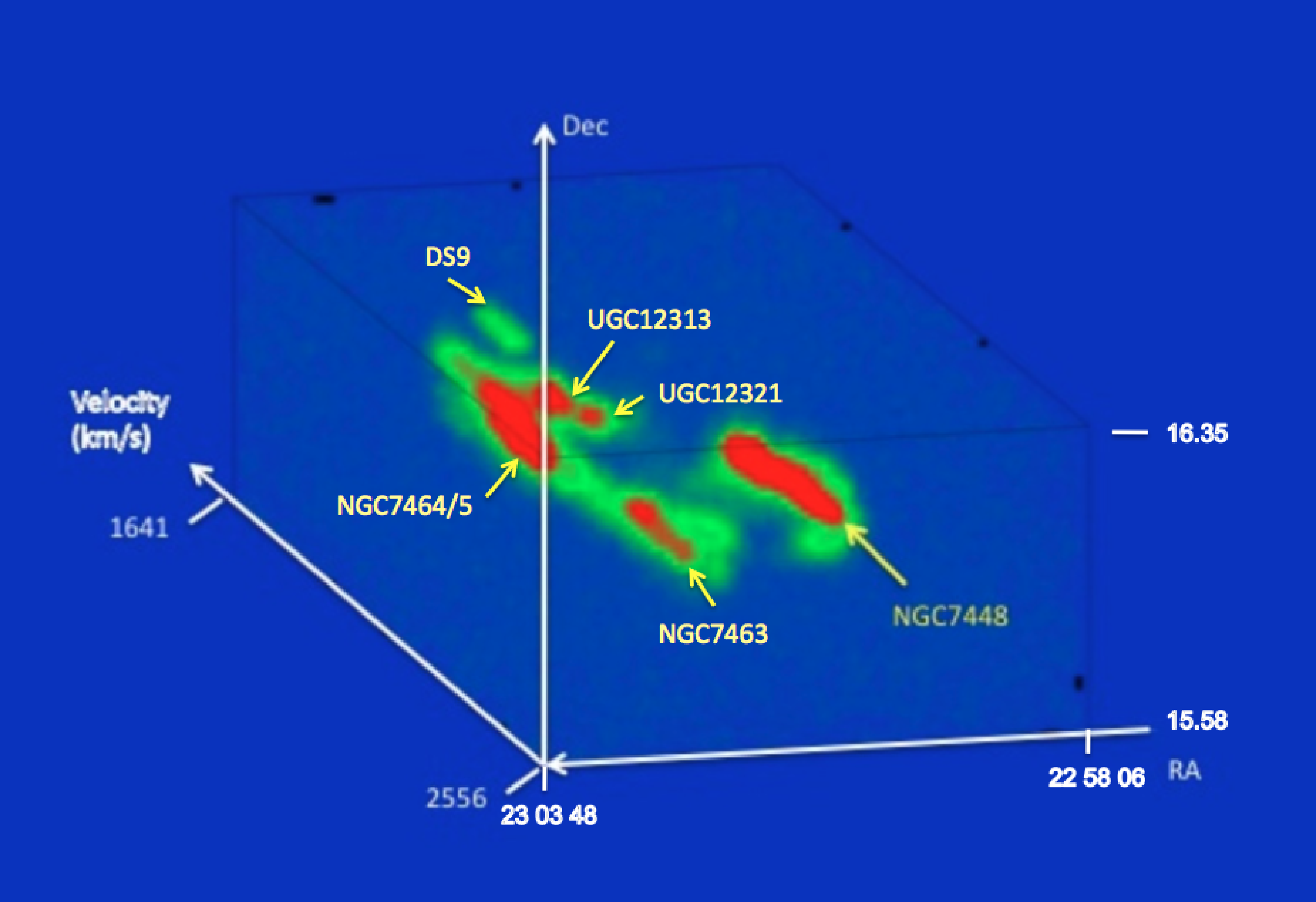}
\caption{We have extracted a region from the HI cube of approximately $1.6\times0.8$ sq deg ($0.8\times0.4$ Mpc at the group distance) extending over a velocity range of 1641 to 2556 km s$^{-1}$. The top image is of the maximum pixel value along the velocity axis to show bright emission irrespective of its velocity. The known galaxies are indicated, along with a stream of HI that extends from the closely grouped galaxies NGC7463/4/5 towards NGC7448, and a tail that protrudes from NGC7448. The Arecibo beam size is indicated by the red circle to the bottom right. The lower image shows the connections and extent of the HI in both spatial and velocity co-ordinates.}
\end{figure}

It is clear that with a 3.5 arc min beam it is difficult to spatially resolve all the galaxies in the group. For example NGC7464 and NGC7465 are separated by about 0.5 arc min and also overlap considerably in velocity with NGC7464 having $V_{HI}\approx1943$ ($\Delta V_{HI}=275$) km s$^{-1}$ and NGC7465 having $V_{HI}\approx1969$ ($\Delta V_{HI}=111$) km s$^{-1}$. Thomas et al. (2002) using the Li and Seaquist (1994) VLA data, suggest that NGC7464 and NGC7465 actually lie within a common HI envelope. To try and overcome this problem we have specifically for these two galaxies used the optical positions of the galaxies and then integrated around this point to obtain the HI parameters given in Table 1. To check our values we can compare them with those obtained at much higher resolution (Westerbork telescope, beam size $0.22 \times 0.85$ sq arc min) by van Driel et al. (1992). The van Driel et al. values are given in brackets in Table 1. Although there are some discrepancies with the measured velocity widths, the integrated fluxes and HI masses agree very well.

The sum of all atomic hydrogen in the six galaxies listed in Table 1 is $1.4 \times 10^{10}$ M$_{\odot}$. It is interesting to see how this mass compares with the mass residing outside of the galaxies, to get some idea about how efficient tidal stripping is in groups like this. To do this we consider the distribution of pixel values throughout the extracted region shown in Fig 3. - this is shown in Fig. 4. The rms value in this smaller data cube is $\approx0.5$ mJy compared with the value of $0.7$ mJy for the whole cube as stated above. This is because this smaller data cube is extracted from the centre of the larger data cube, which becomes noiser closer to its edges. The data closely follow a Gaussian distribution for all negative values and positive values out to approximately 0.2 mJy/pixel, it then flattens out - the Gaussian part is due to the instrumental noise and where it flattens the HI sources (Fig. 4). Fitting a Gaussian to the noise (blue line Fig. 4) gives a more precise measure of the rms of $\approx 0.3$ mJy. This is lower than that quoted above because now the HI sources are not included in the derivation. Subtracting the Gaussian noise (blue line) from the total flux (red line) leaves the flux due to sources in the cube. This leads to a total integrated flux of 287.1 Jy km s$^{-1}$ and a total HI mass of $5\times10^{10}$ M$_{\odot}$. Comparing this to the total contained within the galaxies we find that only about 28\% of the atomic hydrogen in this group is in the galaxies - there is about 2.5 times more atomic hydrogen in the intergalactic medium. This is a similar conclusion to that reached by van Driel et al. (1992). They identify 4 HI clouds within the group and estimate that their mass amounts to 1.5 times that in the galaxies. The van Driel et al. observations were sensitive to atomic gas with column densities more than an order of magnitude higher than ours, which may account for the higher value we find.
\begin{figure}
\centering
\includegraphics[scale=0.52]{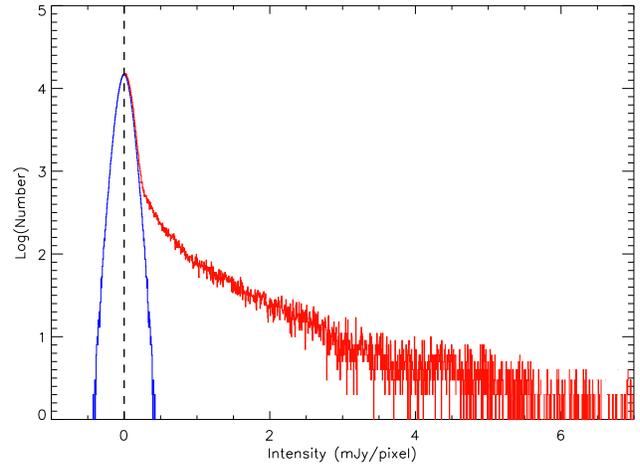}
\caption{Distribution of pixel values for the extracted HI data cube shown in Fig. 3. The blue line shows the Gaussian noise distribution. The red line shows the total intensity (positive values) due to noise and sources.
}
\end{figure}

There are two possible origins of this inter-galactic gas: either it is in-falling because it is left over from the formation phase of the galaxies or it has been drawn out via tidal interactions (or possibly a bit of both). We favour the latter for two reasons. Firstly even though the resolution is poor it does seem that the HI has been drawn out in either a stream or tail (Fig.3) and secondly previous observations (Auld, et al. 2006, Cortese et al. 2008, Minchin et al. 2010, Taylor 2010) indicate that galaxy formation efficiently sweeps up hydrogen into galaxies - there are very few isolated HI clouds with no associated optically detected galaxy (see also Doyle et al. 2005). 
\begin{figure}
\centering
\includegraphics[scale=0.42]{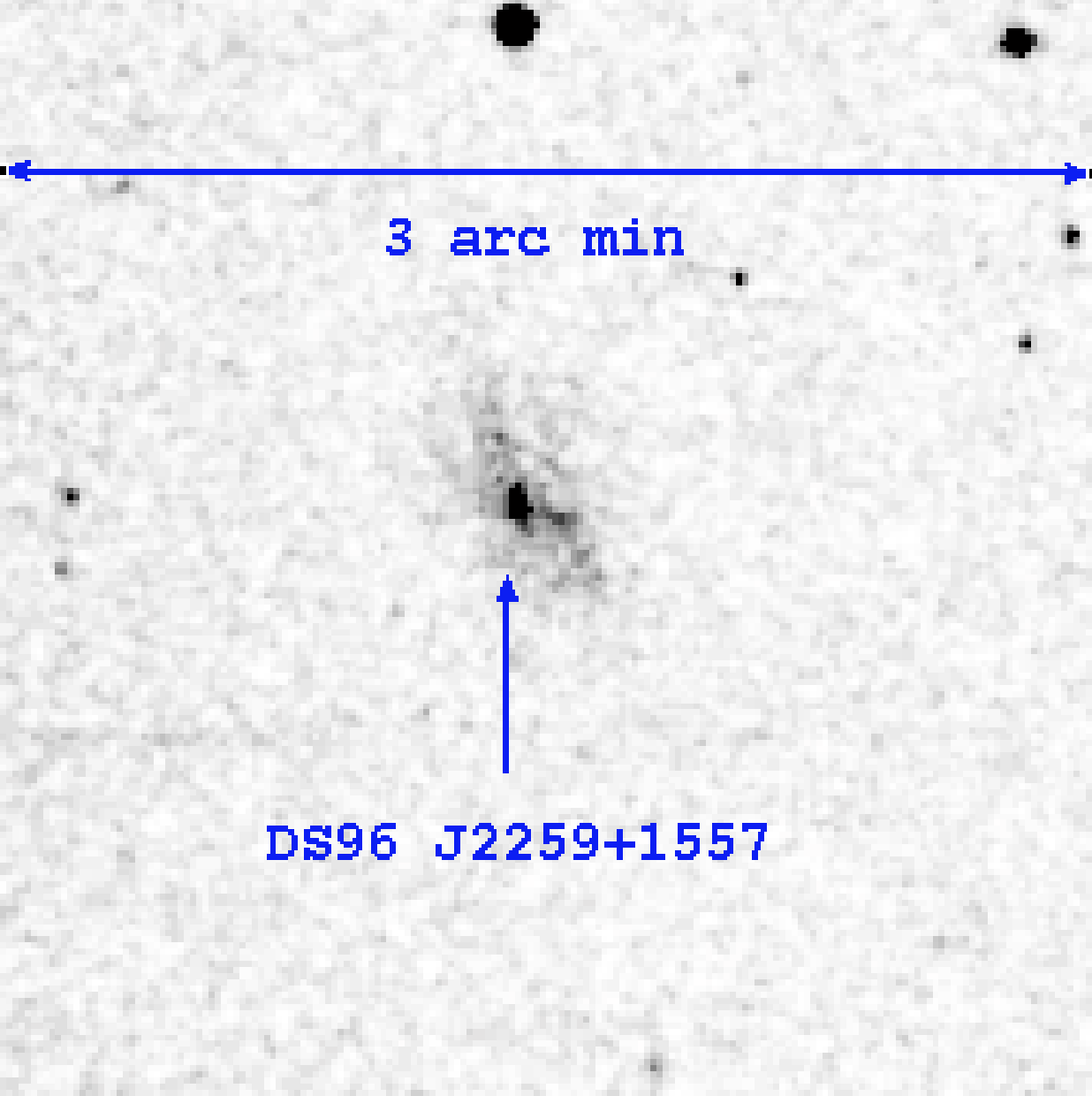}
\includegraphics[scale=0.41,angle=-90]{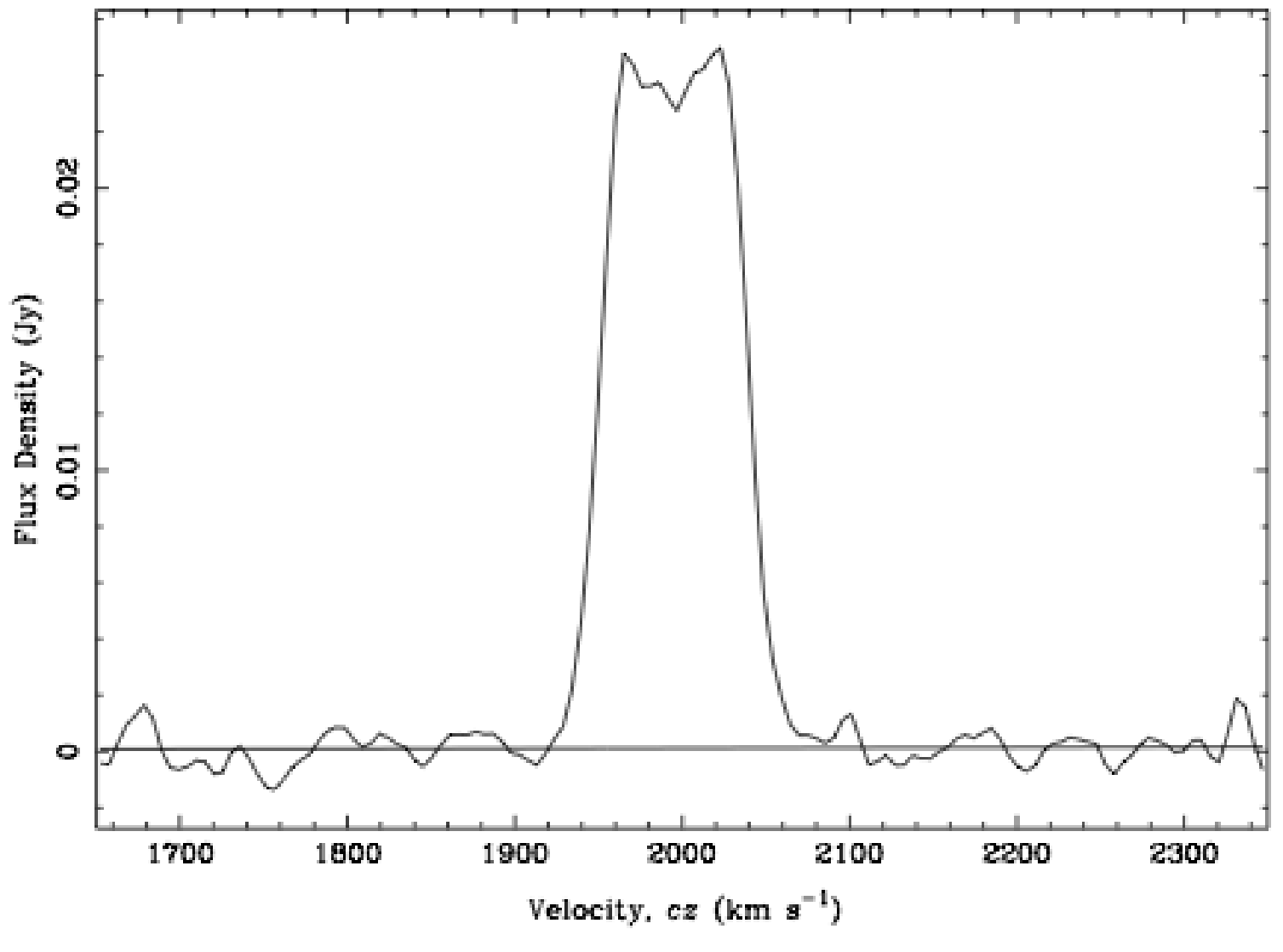}
\caption{The galaxy DS96 J2259+1557. The blue DSS image (above) is centred on the HI position, the HI spectrum is shown below.
}
\end{figure}
\begin{figure}
\centering
\includegraphics[scale=0.42]{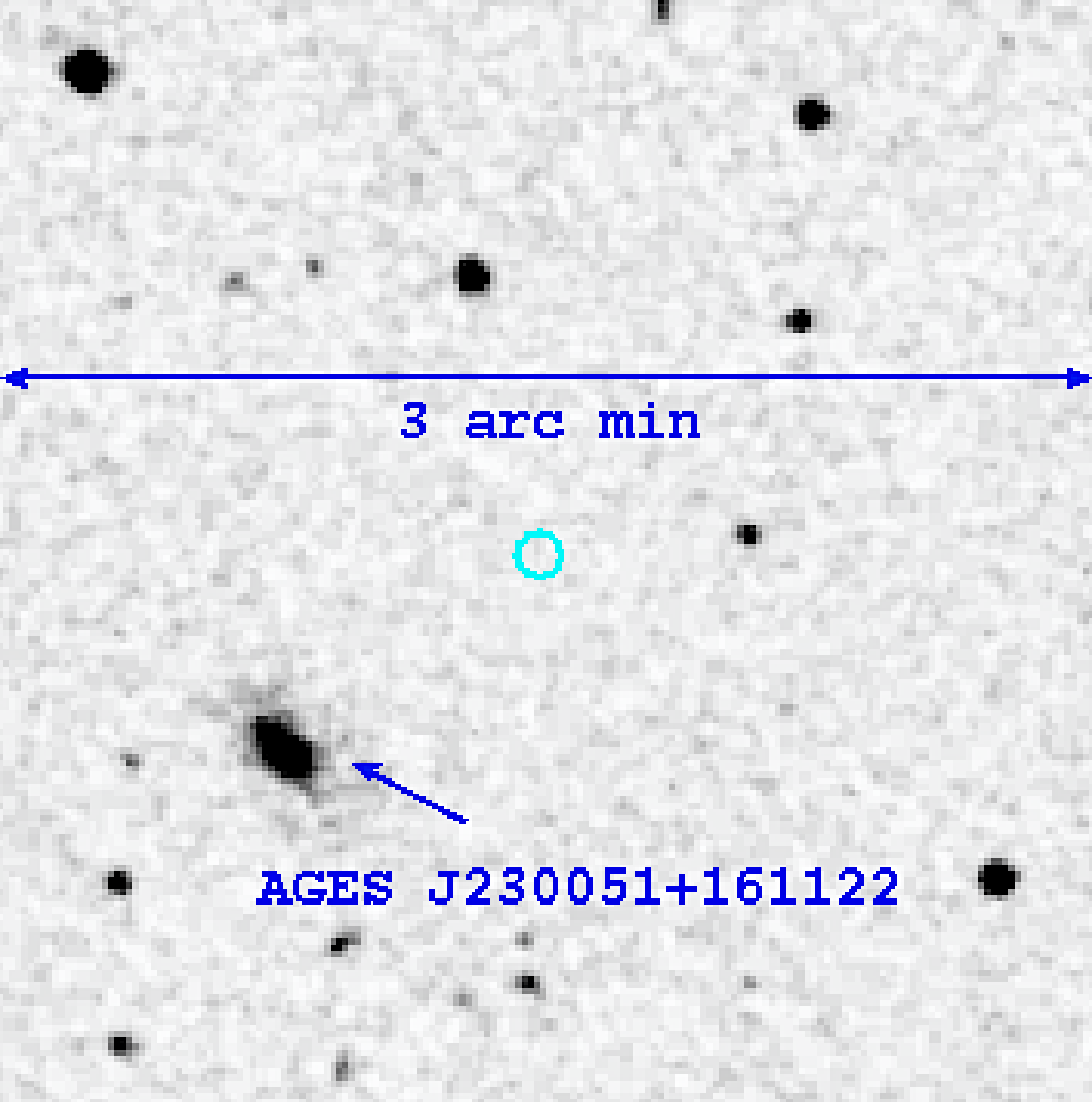}
\includegraphics[scale=0.2]{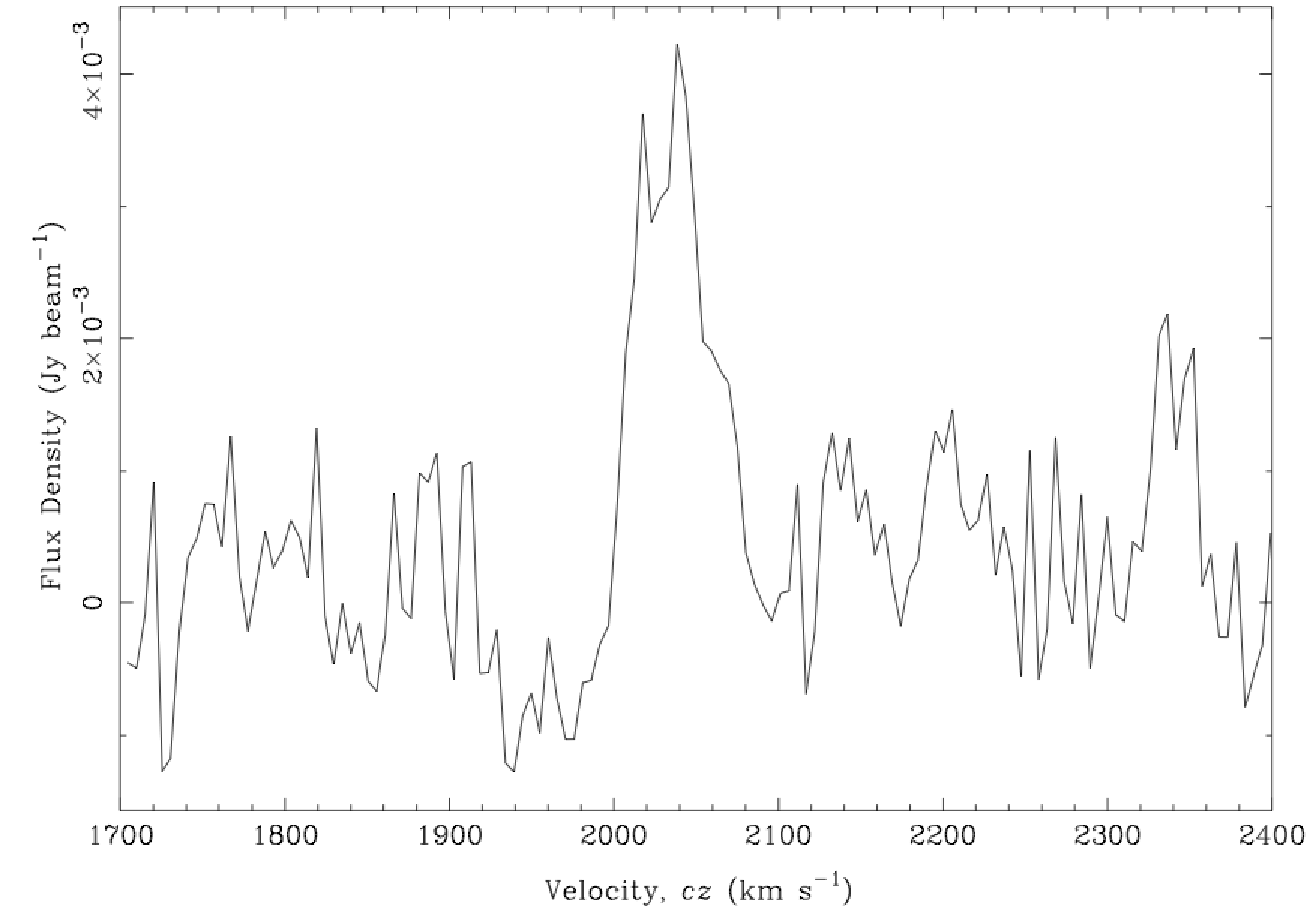}
\includegraphics[scale=0.42]{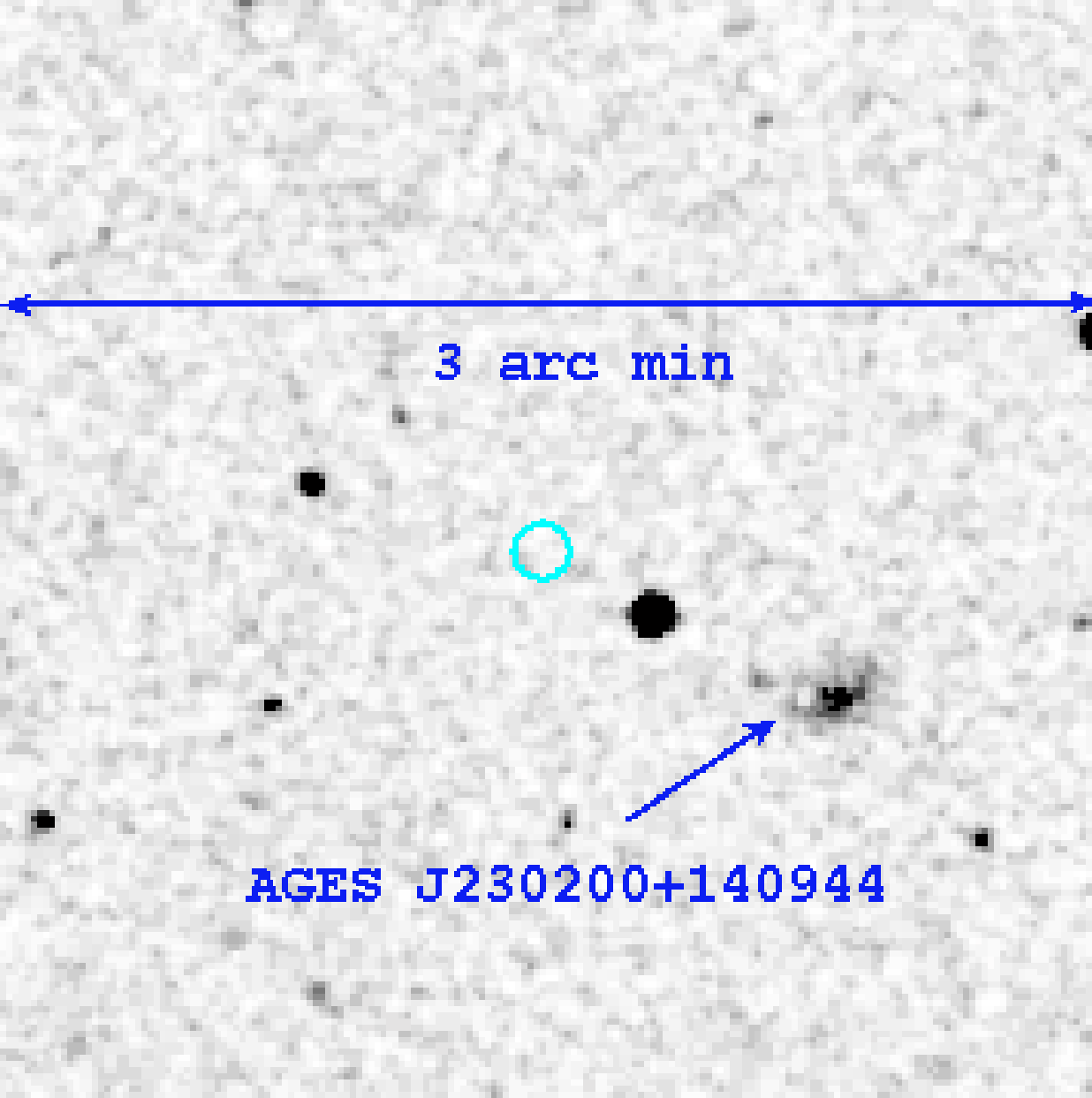}
\includegraphics[scale=0.2]{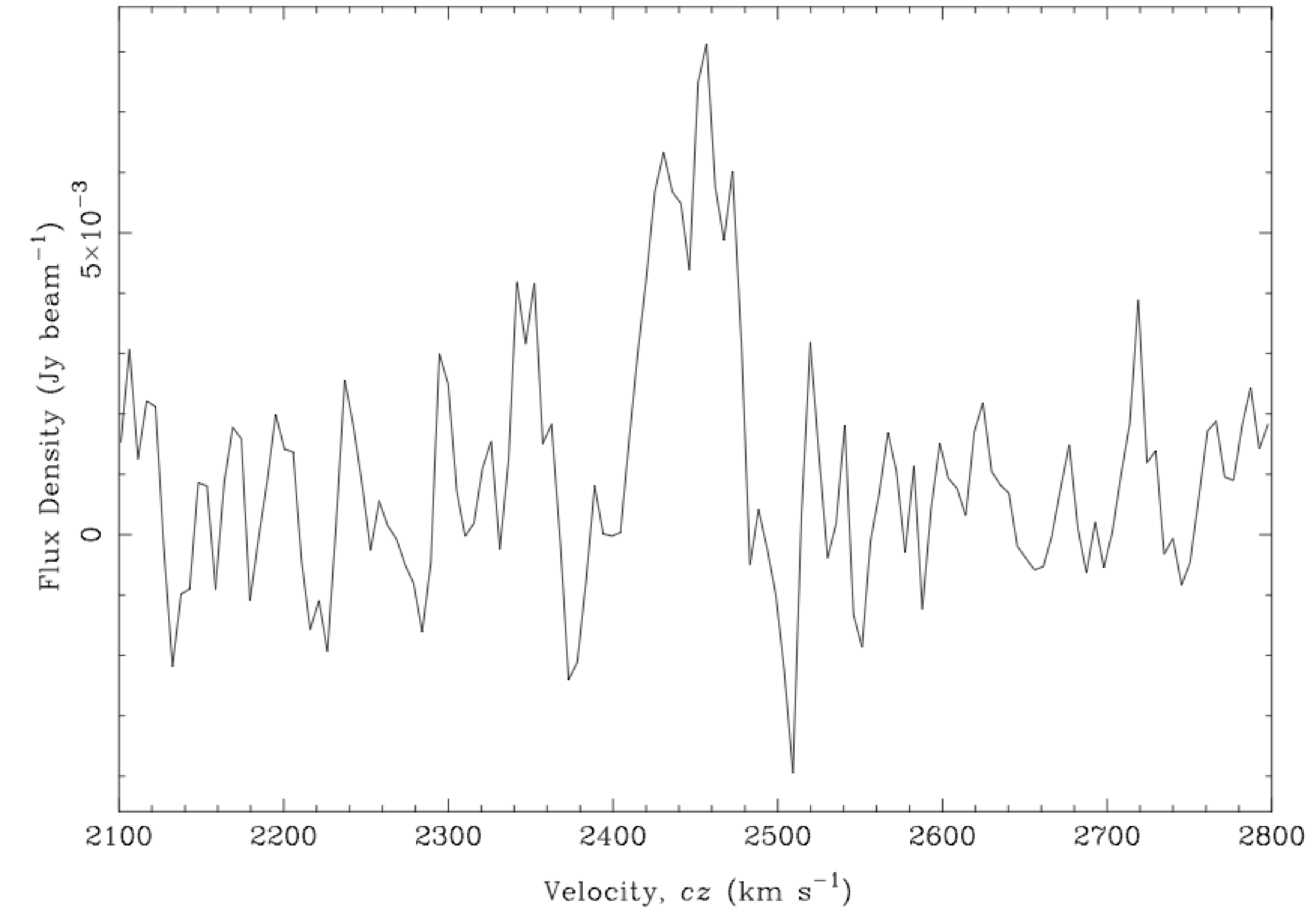}
\caption{Two new dwarf galaxy detections in the NGC7448 group along with their HI spectra. The images are taken from the blue DSS. The light blue circle marks the position derived from the HI observations.
}
\end{figure}

Based on the assumption that these six galaxies in the group are in virial equilibrium we can calculate a group dynamical mass. Following Pisano et al. (2004) we use : \\
\begin{center}
$M_{dyn} = \frac{7.5 \sigma^{2} r_{med}}{G}$ \\
\end{center}
where $\sigma$ is the radial velocity dispersion (155 km s$^{-1}$), $r_{med}$ is half the maximum separation between the galaxies (116 kpc) and $G$ is the gravitational constant. This leads to a dynamical mass of M$_{dyn} = 5 \times 10^{12}$ M$_{\odot}$, which is comparable to the masses found for the six groups studied by Zwaan et al. (2001) - mean value of $7 \times 10^{12}$ M$_{\odot}$. Using the B band magnitudes (corrected for Galactic extinction) given in Li and Seaquist (1994) (see Section 3.1 below and Table 1) we estimate a total stellar mass of $7 \times 10^{10}$ M$_{\odot}$. Combining the stellar and HI mass gives a values of $M_{dyn}/M_{Baryon} < 42$ for the group as a whole. For a typical cosmological model (H$_{0}$=72 km s$^{-1}$, $\Omega_{m}=0.27$, $\Omega_{\Lambda}=0.73$, i.e. Spergel et al. 2007) the ratio of total matter to baryonic matter is $\approx 5$. To be consistent with the cosmic value requires there to be about eight times more baryonic mass in the cold (molecular) and warm/hot (x-ray) gas than in the stars and atomic hydrogen. We note that the NGC7448 group is one of 48 out of 109 nearby galaxy groups that was found not to have diffuse x-ray (ROSAT) emission (Mulchaey et al. 2003) and that typically the warm (x-ray) gas amounts to at most a mass of order that found in the stars (Mulchaey, 2000). This makes the baryon account even more difficult to reconcile. Using the above velocity dispersion and maximum separation the group has a dynamical time of just $7\times10^{8}$ yr, much shorter than a Hubble time. The NGC7448 group is either not in dynamical equilibrium because it has recently formed ($t_{form}<$few$\times10^{9}$ yr) or it has a high dark matter mass fraction.

\begin{table*}
\begin{center}
\begin{tabular}{lccccccccccc}
Name        & RA$_{HI}$  & Dec$_{HI}$ & RA$_{opt}$ & Dec$_{opt}$ & Velocity      & M$_{B}$ & M$^{B}_{star}$  & M$_{HI}$    & M$_{dyn}$   & R$_{iso}$ & $b_{iso}/a_{iso}$\\
            &  (J2000)   & (J2000)    &  (J2000)   & (J2000)     & (km s$^{-1}$) &         & M$_{\odot}$     & M$_{\odot}$ & M$_{\odot}$ & (arc sec) & \\ \hline
AGES J230051+161122 & 23  00 51.2 & 16 11 22          & 23 0 52.0  & 16 10 57           & 2040                   & -14.8   & $1.3\times10^{8}$            & $4\times10^{7}$      & $8\times10^{8}$  & 5.3 & 0.53  \\
AGES J230200+140944 & 23  02 00.4 & 14 09 44          & 23 1 58.8  & 14 09 42            & 2445                   & -13.7   & $4.7\times10^{7}$            & $1\times10^{8}$      & $2\times10^{9}$ & 7.2 & 0.76   \\
\end{tabular}
\caption{Properties of the two newly detected NGC7448 group dwarf galaxies.}
\end{center}
\end{table*}

There is one other central group object that warrants attention. It was originally detected in a HI survey by Duprie and Schneider (1996). The object is listed in NED as DS96 J2259+1557, but we can find no other references to it. DS96 J2259+1557 is closest to UGC12313 both spatially (12 arc min) and in velocity ($\Delta V=24$ km s$^{-1}$). There is no optical counterpart to DS96 J2259+1557 described in the literature, but there is clearly an object at the HI position on the Digital Sky Survey (DSS) image (Fig. 5). There is no photometry for this object - the SuperCOSMOS image (Hambly et al. 2009) breaks this single low surface brightness object up into a number of smaller objects. The HI signal is strong (Fig. 5) with $2 \times 10^{8}$ M$_{\odot}$ of hydrogen rotating at a not corrected for inclination velocity of $W_{50} \approx 89$ km s$^{-1}$ . DS96 J2259+1557 is obviously a dwarf galaxy member of the group and will be an interesting object for further follow up observations.

\subsection{New dwarf galaxies in the NGC7448 group}
During our search of the NGC7448 area we discovered two new dwarf galaxy members of the group (Table 2). AGES J230051+161122 is in a region not covered by the Sloan Digital Sky Survey (SDSS), which only covers about one third of the area observed in HI, but it is clearly visible on the DSS image (Fig. 6) even though there is no entry in NED. AGES J230200+140944 is further to the south in the SDSS area and is identified as SDSS J230158.72+140942.5 with no previous redshift. It is again easily identifiable on the DSS image (Fig 6). AGES J230051+161122 appears to be part of the central sub-group (NGC7463/4/5 and NGC7448) both from its apparent proximity (7.5 arc min from the group centre given in NED) and its velocity (2030 km s$^{-1}$). AGES J230200+140944 is spatially close ($\approx14$ arc min) to UGC12308, but with a higher velocity (2445 compared with 2233 km s$^{-1}$).

Although not in NED AGES J230051+161122 is listed in the SuperCOSMOS catalogue. With a B band magnitude of 17.9 at 28.6 Mpc it has an extinction corrected absolute B magnitude of -14.8. Simply assuming stars like the Sun (M$_{B}$=5.5) this gives a stellar mass of $M_{star}^{B}\approx 1.3 \times 10^{8}$ M$_{\odot}$. Using the standard formula for HI mass: \\
\begin{center}
$M(M_{\odot})_{HI}=2.4 \times 10^{5} d^{2}_{Mpc} F_{Tot} $ \\
\end{center}
where $d_{Mpc}=28.6$ and $F_{Tot}=0.2$ is the integrated HI flux in Jy km s$^{-1}$, we calculate an HI mass of $M_{HI}=4\times10^{7}$ and a value of $M_{HI}/M_{star}^{B}=0.3$. 

We can make an estimate of AGES J230051+161122's dynamical mass (Cortese et al. 2008) using: \\
\begin{center}
$M_{dyn} \approx \frac{R_{iso} W^{2}_{m}}{G}$ \\
\end{center}
where we have assumed that $R_{HI}$ is twice (Salpeter and Hoffman 1996, Swaters et al. 2002) the optical isophotal radius. $W_{m}$ is the inclination corrected mean velocity width (mean of $W_{50}$ and $W_{20}$). Assuming a thin disc $\cos{i}=b_{iso}/a_{iso}$ ($a_{iso}$ and $b_{iso}$ are the SuperCOSMOS semi-major and semi-minor axis respectively) and we use $R_{iso}=\sqrt{a_{iso} b_{iso}}=5.3$ arc sec. This gives a dynamical mass of $M_{dyn} \approx 8 \times 10^{8}$ M$_{\odot}$. Taking M$_{baryon}$=M$_{star}^{B}$+M$_{HI}$ we have a dark matter dominated dwarf galaxy with M$_{dyn}$/M$_{baryon}$$\approx$5.

\begin{table*}
\begin{center}
\begin{tabular}{lccccccccc}
Name        & RA$_{HI}$  & Dec$_{HI}$(J2000) & RA$_{opt}$ & Dec$_{opt}$(J2000) & Velocity (km s$^{-1}$) & M$_{B}$ & M$^{B}_{star}$ M$_{\odot}$ & M$_{HI}$ M$_{\odot}$ & M$_{dyn}$ M$_{\odot}$ \\ \hline
SDSS J230511.15+140345.7 & 23  05 10.8 & 14 04 04          & 23 05 11.1  & 14 03 46            & 1564                   & -15.4   & $2.3\times10^{8}$            & $1\times10^{8}$      & $3\times10^{9}$     \\
SDSS J230615.05+143927.5 & 23  06 14.4 & 14 39 37          & 23 06 15.0  & 14 39 28           & 1542                   & -13.8   & $0.5\times10^{8}$            & $4\times10^{8}$      & $2\times10^{9}$    \\
AGES J230843+161919 & 23  08 43.0 & 16 19 19          & -  & -          & 1610                   & -   & -         & $2\times10^{7}$      & -   \\
\end{tabular}
\caption{Properties of three isolated dwarf galaxies that lie within $V=2104$ km s$^{-1}$.}
\end{center}
\end{table*}

For AGES J230200+140944 we can again use the SuperCOSMOS B band data to get an absolute magnitude of -13.7 and a stellar mass of $M_{star}^{B}\approx4.7\times10^{7}$ M$_{\odot}$. To check the accuracy of this we can in this case use the more extensive SDSS data and the formula from Bell et al. (2003): \\
\begin{center}
$ \log{M_{star}}=-0.222+0.864(g-r)+\frac{M(i)-4.56}{-2.5} $ \\
\end{center}
where $M(i)$ is the $i$ band absolute magnitude, to calculate a stellar mass of M$_{star}=4.0 \times10^{7}$ M$_{\odot}$, very close to that calculated from the B band magnitude. As above we calculate an HI mass of $M_{HI}=1.0\times10^{8}$ M$_{\odot}$ for AGES J230200+140944. This is a gas rich galaxy with $M_{HI}/M_{star}^{B}\approx3$. Even without considering helium, metals and molecular gas this galaxy's baryonic mass is apparently dominated by gas and not stars, making it young or slowly evolving. With $(g-r)=0.4$
 AGES J230200+140944 is blue compared to the mean value for SDSS galaxies of $(g-r) \approx 0.8$, $(g-r)$ has a range of approximately 0.3-1.0 (Bell et al. 2003). Again as above we calculate a dynamical mass of $M_{dyn} \approx 2\times 10^{9}$ M$_{\odot}$ (in this case $R_{iso}=\sqrt{a_{iso} b_{iso}}=7.2$ arc sec). Taking M$_{baryon}$=M$_{star}$+M$_{HI}$ we have a dark matter dominated galaxy with M$_{dyn}$/M$_{baryon}$$\approx$15.  AGES J230051+161122 and AGES J230200+140944 are very similar in total masses, but AGES J230200+140944 has a much higher fractional gas and dark matter content.

\section{New dwarf galaxies external to the NGC7448 group}
One of the stated aims of AGES is to try and discover gas rich dwarf galaxies that are not detected in optical surveys. To specifically look for dwarf galaxies not associated with the NGC7448 group we have also considered HI sources that lie within the nearby Universe i.e. less than the group velocity of $V = 2104$ km s$^{-1}$. We identify three sources and give their properties in Table 3. Two of the sources are in the area covered by the SDSS and can be identified with SDSS sources. Both SDSS J230511.15+140345.7 and SDSS J230615.05+143927.5 lie about one degree away (500 kpc at 28.6 Mpc) from the nearest group galaxy UGC12308 and are at a velocity some 700 km s$^{-1}$ lower. They are separated from each other by a little over 30 arc min and appear to be isolated from larger galaxies.  The third galaxy (AGES J230843+161919) has the lowest HI mass in the sample. It is not in the area covered by SDSS and there is no object listed in NED. Unlike the other two galaxies (see below) there is no clear object to be seen on the DSS either, though there is a very faint smudge that requires investigating with much deeper observations.  AGES J230843+161919 also seems to be isolated from the brighter galaxies in this region of sky. It is more than 1.5 degrees (750 kpc at 28.6 Mpc) away from any of the bright group galaxies and has a velocity about 300 km s$^{-1}$ lower. With no optical identification we cannot calculate a stellar mass.

SDSS J230511.15+140345.7 is listed as a star in the SDDS database, but the position sits right at the centre of what is most certainly a galaxy. There is also a spectrum which contradictorily is described as a galaxy rather than a stellar spectrum. The optical spectrum is clearly that of a star forming galaxy with a blue continuum and emission lines. From the optical spectrum a velocity of 1530 km s$^{-1}$ is measured which compares with a value of 1564 km s$^{-1}$ from the HI. SDSS give a value of g=22.94 for this source which is clearly too faint and is probably to do with the stellar mis-classification. Using the SuperCOSMOS data as described above we find a value of m$_{B}$=16.7 and a Galactic extinction corrected absolute magnitude of M$_{B}=-15.4$. Using the absolute magnitude of the Sun as above, this equates to a stellar mass of $2.3 \times 10^{8}$ and a value of $M_{HI}/M_{Star}^{B}=0.4$. The SuperCOSMOS data also provides us with an isophotal size and an ellipticity. Using the same method as described above leads to a dynamical mass of $3 \times 10^{9}$ M$_{\odot}$, about a factor of 10 higher than the combined baryonic mass in stars and atomic gas.

Our HI position is just 0.2 arc min away from that of the optical position of SDSS J230615.05+143927.5. The SDSS image shows SDSS J230615.05+143927.5 to be a very low surface brightness object. There is a small puzzle in that there is also an SDSS spectrum which gives a velocity of just 1233 km s$^{-1}$ some 300 km s$^{-1}$ less than the HI velocity, but because the galaxy is of very low surface brightness the SDSS spectrum is very noisy with no clear emission lines and so we suggest that the SDSS velocity is incorrect. Going through the same calculations described above we get M$_{B}=-13.8$, a stellar mass of $0.5 \times 10^{8}$ and a very large $M_{HI}/M_{Star}^{B}=8$. The baryonic mass of this galaxy seems to be totally dominated by gas rather than stars. To put this in perspective Minchin et al. (2010) recently calculated $M_{HI}/M_{Star}^{B}$ values for 87 galaxies selected from the AGES NGC1156 and NGC7332 fields. They found a mean value of $M_{HI}/L_{B}$ for these 87 galaxies of $1.6 \pm 0.2$ with lowest and highest values of 0.14 and 11.3 respectively. We note that for low surface brightness galaxies like SDSS J230615.05+143927.5 SuperCOSMOS may underestimate magnitudes in which case $M_{HI}/M_{Star}^{B}$ would be smaller. As above we calculate a dynamical mass of $2 \times 10^{9}$ M$_{\odot}$, about a factor of 4 higher than the combined baryonic mass in stars and atomic gas.

\section{Galaxies detected in the NGC7448 volume}

\subsection{Source extraction}
A crucial part of deriving a well defined sample of HI sources from a blind HI survey is source extraction with defined selection criteria. 
Three of us independently searched the HI data cube by eye, resulting in 176, 151 and 141 detections. By combining these lists we have a provisional list of 191 objects detected by one, two or three of us. 52 objects were detected by just one of us, 139 by at least two of us and 92 by all three of us. For each detection we again used the MBSPECT routine to fit the data to obtain, recessional velocity, peak flux, total flux, and velocity width at 20 and 50\% peak height. Fig. 7 is a plot of total 21cm flux against the line width at 50\% peak flux. All detections are shown with those detected by at least two of us marked in blue and those by just one in red. To compare with other AGES papers (i.e. Cortese et al. 2008) we also draw a line on Fig. 7 where the signal-to-noise ratio is equal to 6.5 (Saintonge, 2007) where: \\
\begin{center}
$S/N_{Tot}=\frac{1000 \times F_{Tot}}{W_{50}} \times \frac{w^{1/2}}{\sigma}$ \\
\end{center}
$F_{Tot}$ is the total flux in Jy km s$^{-1}$, $W_{50}$ is the velocity width at 50\% peak height, $w$ is either $W_{50}/(2 \times \Delta V)$ for $W_{50} < 400$ km s$^{-1}$ or $400/(2 \times \Delta V)$ for $W_{50} \ge 400$ km s$^{-1}$, $\Delta V$ is the velocity resolution (10 km s$^{-1}$) and $\sigma$ the rms noise in mJy. Saintonge (2007) found that the vast majority (95\%) of sure sources lie above the $S/N_{Tot}=6.5$ line (see also Cortese et al. 2008). It is clear from Fig. 7 that almost all of the 139 detections confirmed by two people lie above the $S/N_{Tot}=6.5$ line. 
\begin{figure}
\centering
\includegraphics[scale=0.5]{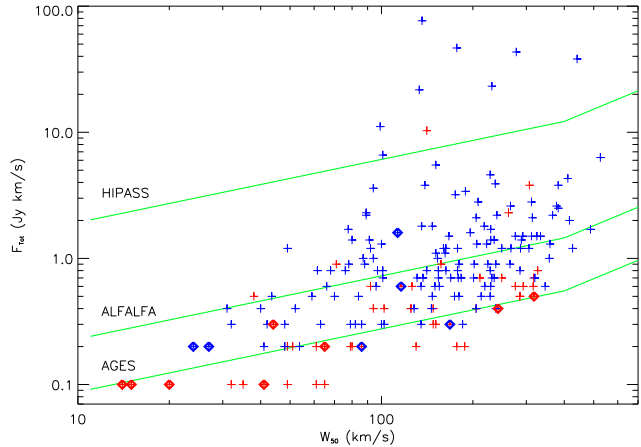}
\caption{AGES total flux against the velocity width at 50\% peak intensity. The complete sample of 191 detections is shown with those identified independently by at least two of us (139) marked in blue, those by just one of us in red. Those that have been subject to follow up observation are marked with a diamond - blue a confirmed detection, red a non-detection. The green lines indicate, from top to bottom, the $S/N_{Tot}=6.5$ line for the HIPASS, ALFALFA and AGES surveys respectively.
}
\end{figure}

We have carried out follow up observations at Arecibo of 14 sources detected by just one of us - these are marked with a diamond on Fig. 7 - red for a subsequent non-detection, blue for a confirmed detection. Only 2 out of 9 sources followed up with $S/N_{Tot}<6.5$ have been confirmed, while 4 out of five have been confirmed with $S/N_{Tot}>6.5$, indicating that a small fraction of our sources may turn out to be false on follow up.  Based on the above we define our HI sample to be all those galaxies with $S/N_{Tot}>6.5$, 175 galaxies. From the information given in NED 142 of the 175 detections have no previous HI measurements. The Hubble velocity distribution for the complete sample (175) is shown in Fig. 8. There is clear evidence for large scale structure with prominent peaks at about 2000, 7,000, 9,000 and 11,000 km s$^{-1}$.

For comparison we have also included on Fig. 7 the $S/N_{Tot}=6.5$ line for two other recent blind large area HI surveys - HIPASS (Meyer et al., 2004)) and ALFALFA (Giovanelli et al., 2005, Giovanelli et al. 2007). These surveys are characterised by having approximate values of $\sigma=13$ and 2.5 mJy respectively. The lines illustrate the increasing numbers of low total flux and small velocity width galaxies that are detected as the integration time increases. For example only 9 of these galaxies would have $S/N_{Tot}>6.5$ in the HIPASS survey and 73 in the ALFALFA survey compared to our 175. The HIPASS and ALFALFA surveys in total detect many more galaxies than AGES because they cover much larger areas, but within a given volume AGES detects many more galaxies than HIPASS and ALFALFA.
\begin{figure}
\centering
\includegraphics[scale=0.51]{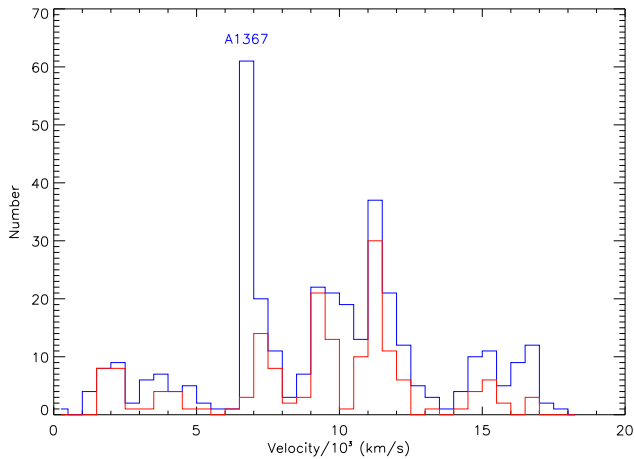}
\caption{The distribution of Hubble velocities for all 175 HI detections in the NGC7448 data cube with $S/N_{Tot}>6.5$ is shown in red. The blue line shows the velocity distribution of the complete AGES sample, which includes observations of the cluster A1367.
}
\end{figure}

\subsection{HI mass}
As above we have calculated a HI mass for all 175 detections,
$d_{Mpc}$ is calculated from the measured Hubble velocity using H$_{0}$=72 km s$^{-1}$ Mpc$^{-1}$. The distribution of HI mass is shown in Fig 9a. The lowest detected HI mass object has M$_{HI}=2 \times 10^{7}$ M$_{\odot}$ and lies at a velocity of 1610 km s$^{-1}$, it was discussed in detail in section 3.1. The highest HI mass is M$_{HI}=3 \times 10^{10}$ M$_{\odot}$ and can clearly be identified with UGC12374 because the optical (NED) and HI position differ by only 0.2 arc min and the velocities by only 12 km s$^{-1}$. In NED it is classified as an SBb galaxy, but is otherwise undistinguished as there are only references to it appearing in catalogues. Only a small part of the HI data cube is covered by the SDSS, but UGC12374 is in this region so we can again use the prescription of Bell et al. (2003) to calculate a stellar mass of $M_{Star}=2.5 \times 10^{11}$ M$_{\odot}$. This is a galaxy that has a massive amount of HI, but its HI to stellar mass ratio is not unusual at about 0.1. UGC12374 has just about the same HI mass as the massive HI galaxy AGES J224005+244154 discussed in detail in Minchin et al., (2010), but unlike UGC12374 the baryonic mass of AGES J224005+244154 is dominated by the gas and not the stars. 

Correcting for the volume sampled we can convert the HI mass distribution into a HI mass function. We use the $1/V_{max}$ method (Felton, 1977) where $V_{max}$ is the volume a galaxy can occupy given its HI mass, velocity width ($W_{50}$) and a signal-to-noise ratio of $S/N_{Tot}=6.5$. We note that the $1/V_{max}$ method can be sensitive to large scale structure, but it does have the advantage over other methods that there is no a priori assumptions about the form of the mass function and whether it changes shape within different environments. For the similarly derived luminosity function Cole et al. (2001) find that luminosity functions derived using the $1/V_{max}$ and maximum likelihood methods (as used by Martin et al. 2010, see below) are identical within the errors.
To improve the accuracy of our mass function we have added to our NGC7448 data the data obtained from other regions of the sky observed as part of AGES (the AGES volume sample). The additional galaxies come from fields centred on NGC628 (Auld et al. 2006), NGC1156 and NGC7332 (Minchin et al. 2010) and A1367 (Cortese et al. 2008). Using the same criteria ($S/N_{Tot}>6.5$) for this additional data we now have a sample of 370 galaxies. The distribution of Hubble velocities for the complete sample is shown in Fig. 8. \footnote{Following Cortese et al. (2008), for galaxies in the A1367 sample, we use D=92.8 Mpc to calculate HI masses if they have velocities between 4000 and 9000 km s$^{-1}$.} In order to assess the completeness of our data i.e. do the galaxies in our sample fill the volume we expect them to, we have carried out a $V/V_{max}$ test.  We find a mean value of $<V/V_{max}>=0.5$ (0.4969) consistent with galaxies that uniformly fill the volume sampled. 

A Schechter fit to the mass function (Fig. 9b) gives the following parameters $\alpha=-1.52(\pm 0.05)$, $M^{*}=5.1 (\pm 0.3) \times 10^{9}$ M$_{\odot}$, $\phi=8.6 (\pm 1.1) \times 10^{-3}$ Mpc$^{-3}$ dex$^{-1}$. In Table 4 and Fig. 9b we compare these values with those derived by the two major wide, but shallow surveys, of Zwaan et al. (2005) using 4,315 galaxies detected in the HIPASS and the ALFALFA survey of Martin et al. (2010) using 10,119 galaxies. Our faint end slope is significantly steeper than both Zwaan et al. and Martin et al., our value for M$^{*}$ is lower and $\phi$ is larger than the other two surveys. Integrating our mass function leads to a local atomic gas cosmic density of $7.9\times10^{7}$  M$_{\odot}$ Mpc$^{-3}$ a factor of 1.7 higher than found by Zwaan et al., but consistent within the errors with the value found by Martin et al. Our deeper HI survey ($\sigma \approx 0.7$ mJy) has revealed proportionally more low mass galaxies than that derived from the HIPASS survey ($\sigma \approx 13$ mJy) and the ALFALFA survey ($\sigma \approx 2.5$ mJy). For a typical cosmological model (H$_{0}$=72 km s$^{-1}$, $\Omega_{m}=0.27$, $\Omega_{\Lambda}=0.73$) the closure density is $\approx1.5 \times10^{11}$ M$_{\odot}$ Mpc$^{-3}$, giving $\Omega_{HI}=5.3(\pm0.8) \times 10^{-4}$. The HI gas is but a small fraction of the total mass/energy (0.05\%), matter (0.2\%) and baryonic matter (1.3\%). 

Martin et al. (2010) provide an extensive discussion on biases that can affect the derivation of the HI mass function. Errors in distance arise from using Hubble's law, peculiar velocities affect nearby, and hence low mass galaxies proportionally more. The mass function low mass slope may also become steeper due to the presence of local large scale structure, which provides more low mass galaxies than is typical for other regions of the sky. Martin et al. consider a number of adjustments to the mass function determination to try and account for these problems. They have a velocity field model that makes adjustments to galaxy distances and a weighting scheme dependent on galaxy density for the $1/V_{max}$ method. They also use an alternative maximum likelihood method to obtain the mass function. The ALFALFA mass function values given in Table 4 are derived using this maximum likelihood method and surprisingly give a faint end slope steeper than that obtained by the $1/V_{max}$ method (-1.33 compared to -1.25). All these methods are used to try and obtain 'global' parameters from data that cover a wide range of galaxy environments, in particular the ALFALFA data set includes part of the nearby Virgo cluster and local supercluster. 

An important question to ask is why does AGES find a steeper mass function faint end slope than other HI surveys such as ALFALFA? We suggest two alternative explanations. The first is that ALFALFA have missed many of the low mass galaxies that AGES detects. For example if we consider the five dwarf galaxies discussed in section 3 and 4 only one of these has a high enough signal-to-noise ratio to be detected by the ALFALFA survey. If all volumes of the nearby Universe were like this then $\sim$80\% of the HI would be missed by ALFALFA in this dwarf galaxy mass range. The rational for the ALFALFA survey was that for uniformly distributed galaxies there would be other volumes where these galaxies would be closer to us and so they would be found and correctly accounted for, which is just the reason for choosing a large area rather than a deep survey. This argument breaks down though if the galaxies are not uniformly distributed around us (cosmic variance). For example the nearest detected galaxy to us in the NGC7448 volume lies at 1610 km s$^{-1}$ (22.4 Mpc). If all galaxies were at or beyond this distance then ALFALFA would detect nothing below about $M_{HI}\approx10^{7.6} $M$_{\odot}$, the only way to find lower mass galaxies would be to make a deeper survey like AGES. Martin et al. (2010) specifically mention the huge void in part of their data, in front of the Pisces-Perseus supercluster, which extends out to $\sim$3000 km s$^{-1}$. So it is possible that ALFALFA is missing low mass galaxies because of the nature of the nearby large scale structure. We note that Rosenberg and Schneider (2002) loosely sampling the Arecibo sky to a similar depth as ALFALFA, but excluding the Virgo cluster, derive the same low mass slope to the HI mass function as AGES ($\alpha=-1.53$, $M^{*}=7.6\times 10^{9}$ M$_{\odot}$, $\phi=5.0\times 10^{-3}$ Mpc$^{-3}$ dex$^{-1}$, $\Omega_{HI}=4.6\times 10^{-4}$).

The second explanation is that AGES is finding too many low mass galaxies and a higher density because of the large scale structure. All our fields are centred on well known galaxies and although we showed in section 3 that three of our five nearby dwarfs do not appear to be associated with the NGC7448 group, they may still be associated with the large scale structure that the group resides within. In this way our survey maybe biased and not typical of the Universe as a whole - you need to include both voids and galaxy structures in a luminosity function calculation and a determination of $\Omega_{HI}$. To try and assess the influence of the foreground structures we have created a mass function with them removed. We exclude galaxies from the NGC1156 field with $200<v<600$ km s$^{-1}$, NGC628 field with $500<v<800$ km s$^{-1}$, NGC7332 field with $1100<v<1350$ km s$^{-1}$, NGC7448 field with $1850<v<2350$ km s$^{-1}$ and A1367 field with $4000<v<9000$ km s$^{-1}$ (302 galaxies remaining). This leads to a mass function with the following parameters $\alpha=-1.47(\pm 0.06)$, $M^{*}=5.6 (\pm 0.4) \times 10^{9}$ M$_{\odot}$, $\phi=6.2 (\pm 0.9) \times 10^{-3}$ Mpc$^{-3}$ dex$^{-1}$ and $\Omega_{HI}=4.2(\pm0.8) \times 10^{-4}$ (Table 4). Although the slope is not so steep the removal of the foreground galaxies does not significantly alter our results.

Our fields are centred on isolated galaxies, groups or clusters, but with the exception of A1367 none are in the supergalactic plane: NGC628 was chosen as an isolated group (with NGC660, UGC1195 and UGC1176) as was NGC7448. NGC7332 and 7339 are an isolated galaxy pair and NGC1156 is an isolated galaxy with no bright companions. NGC7332 and NGC7448 are actually in the local void.  Although A1367 is too far away for low mass galaxies ($M_{HI}<10^{8.5}$ M$_{\odot}$) to be  detected, and so greatly affect the faint end slope of the mass function, we have also considered the result of excluding just the A1367 (super galactic plane) galaxies from our determination of the mass function, the result is $\alpha=-1.52(\pm 0.06)$, $M^{*}=5.5 9\pm 0.4) \times 10^{9}$ M$_{\odot}$, $\phi=6.5 (\pm 0.8) \times 10^{-3}$ Mpc$^{-3}$ dex$^{-1}$ and $\Omega_{HI}=4.3(\pm0.8) \times 10^{-4}$ (Table 4). Excluding the A1367 galaxies does not greatly alter our original conclusions with regard to a relatively steep faint end slope. A better insight into these differences in faint end slope will probably come from a more comprehensive study of the spatial distribution of low HI mass galaxies in surveys like ALFALFA.
\begin{figure}
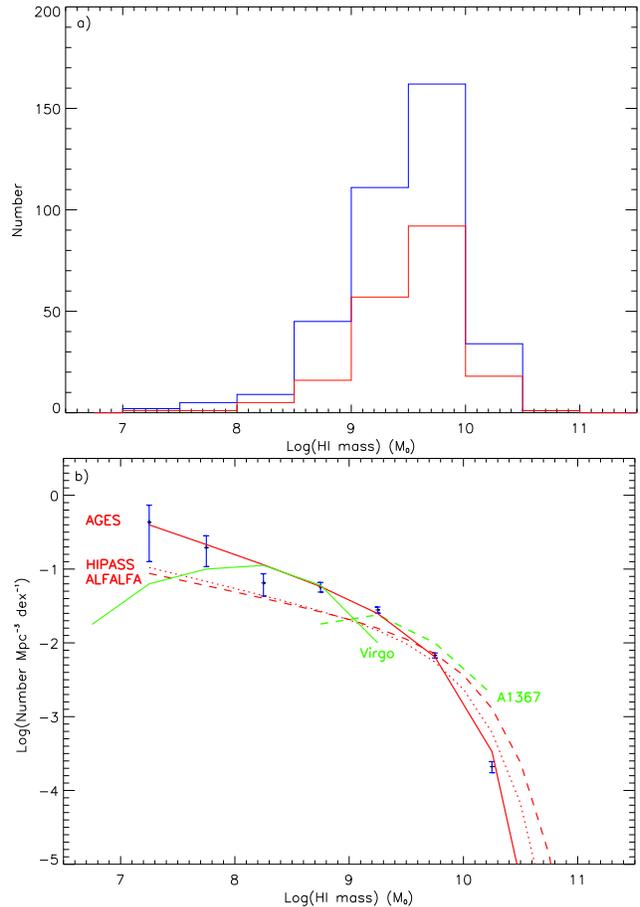

\centering
\includegraphics[scale=0.51]{mass_dist.epsi}
\includegraphics[scale=0.51]{mass_func.epsi}
\caption{a) The distribution of HI mass. The red line is for all 175 galaxies in the NGC7448 sample, the blue line is the complete AGES sample. b) The measured HI mass function (blue points) for the AGES sample, the solid red line is a Schechter law fit to the data. The red dotted and red dashed lines are the Schechter law mass functions from the HIPASS and ALFALFA surveys respectively. The two green lines show the mass function for two clusters A1367 and Virgo, these are arbitrarily normalised at their peak value.
}
\end{figure}
\begin{table*}
\begin{center}
\begin{tabular}{lccccc}
Survey       & $\sigma$ (mJy)  & $\alpha$ & M$^{*}$ (M$_{\odot}$) & $\phi$ (Mpc$^{-3}$) & $\Omega_{HI}$  \\ \hline
AGES & 0.7 &  -1.52$(\pm0.05)$  &   $5.1(\pm0.3) \times 10^{9}$    & $8.6(\pm1.1) \times 10^{-3}$     & $5.3(\pm0.8) \times 10^{-4}$    \\
AGES(-GC) & 0.7 & -1.47$(\pm0.06)$ &  $5.6(\pm0.4) \times 10^{9}$    & $6.2(\pm0.9) \times 10^{-3}$     & $4.2(\pm0.8) \times 10^{-4}$    \\
AGES(-A1367) & 0.7 & -1.52$(\pm0.06)$ &  $5.5(\pm0.4) \times 10^{9}$    & 
$6.5(\pm0.8) \times 10^{-3}$     & $4.3(\pm0.8) \times 10^{-4}$    \\ 
ALFALFA & 2.5 &   -1.33$(\pm0.02)$  &   $9.1(\pm0.4) \times 10^{9}$  & $4.8(\pm0.3) \times 10^{-3}$     & $4.4(\pm0.3) \times 10^{-4}$    \\
HIPASS & 13.0 &   -1.37$(\pm0.03)$  &   $6.8(\pm0.5) \times 10^{9}$   & $5.0(\pm0.7) \times 10^{-3}$     & $3.2(\pm0.5) \times 10^{-4}$    \\
\end{tabular}
\caption{A comparison of Schechter fit HI mass function parameters obtained by large area blind surveys. AGES parameters are for the whole sample, AGES(-GC) are the parameters excluding galaxies in the foreground groups and clusters, AGES(-A1367) excluding just the A1367 galaxies. HIPASS comes from Zwaan et el. (2005) and ALFALFA from Martin et al. (2010). All values have been calculated for a typical cosmology (H$_{0}$=72 km s$^{-1}$, $\Omega_{m}=0.27$, $\Omega_{\Lambda}=0.73$).}
\end{center}
\end{table*}

We have already noted how the environment can affect the way a galaxy evolves with time. We have too few group galaxies to compare our global HI mass function with that of galaxy groups, but we can compare with a galaxy cluster. Also as part of the AGES Taylor (2010) has produced a HI mass function (Fig 9b) for two regions of the Virgo cluster covering 15 sq deg. At face value the Virgo cluster mass function lacks both the high and low mass galaxies that are found elsewhere, though at the low mass end the error bars are large. Although covering different regions of the Virgo cluster the Taylor result agrees with Davies et al. (2004) who also found that the Virgo cluster HI mass function lacked both high and low HI mass galaxies when compared to the global HI mass function. Both the Taylor and Davies et al. Virgo HI mass functions peak at the same point $\log{M_{HI}} \approx 8.25$. It is easy to understand how low mass and typically low surface density galaxies may lose their gas in the cluster environment via tidal effects, accelerated star formation and ram pressure stripping. One might expect that high mass galaxies are more robust. A flatter low mass slope compared to the 'global' value has been confirmed by surveys of other moderately dense environments such as the Ursa Major cluster (Verheijen et al. 2000) and the Canes Venatici region (Kovac et al. 2009).

We have also derived a mass function from the A1367 data (54 galaxies) which is in reasonably good agreement with the global HI mass function, though at a distance of 92.8 Mpc this does not extend to very low HI masses (Fig 9b). The A1367 data covers not only the central regions of the cluster, but also a substantial part of the outer infalling region (Cortese et al. 2008).
By comparing Virgo and A1367 we may glean some idea about what is going on. A1367 is currently being assembled from infalling galaxy groups (Cortese et al. 2004), thus explaining  why its HI mass function is similar to that of the field over this mass range. Cortese et al. (2008) also identify a number of bright A1367 galaxies that are surrounded by diffuse hydrogen, an indication that gas is being lost from more massive galaxies as they fall into the cluster potential, but this process has not proceeded as far as it has in Virgo where our observations exclusively cover the more central regions. If these A1367 infalling groups are like the NGC7448 group then we already know from our discussion above that there could be as much as 2.5 times more HI outside the galaxies than inside, and this is before the cluster has exerted its influence. We conclude that the different shape of the Virgo cluster HI mass function (Taylor 2010, Davies et al. 2004) compared to that of the general field is the natural result of environmental influences. HI deficiency in Virgo cluster galaxies has previously been well documented (e.g. Haynes et al. 1984).

As stated above one of the many objectives of the AGES was to see if the numbers of detected dwarf galaxies around nearby large galaxies would increase when we searched using a deep HI rather than an optical survey. If theory and observation are to agree then many more dwarf galaxies per unit volume need to be discovered (Klypin et al. 1999). If the HI mass function directly followed the predicted dark matter mass function with power law slope of $\alpha \approx -2$ we should expect to find $\approx 10$ galaxies with $M_{HI} \approx 10^{8}$ M$_{\odot}$ for each galaxy with $M_{HI} \approx 10^{9}$ M$_{\odot}$. If the galaxy group HI mass function follows the same power law as that derived above for the AGES sample then we would expect about 4 galaxies with $10^{8}$ M$_{\odot}$ for each galaxy of mass $M_{HI} \approx 10^{9}$ M$_{\odot}$. This is clearly incompatible with our observations as there should be roughly 24 companions to the six central group galaxies. In this paper we have found just two additional dwarf galaxies of masses of order $10^{8}$ M$_{\odot}$ in a galaxy group containing six galaxies with masses of order $10^{9}$ M$_{\odot}$ (Tables 1 and 2). This result is consistent with our previous findings. We have found two dwarf galaxies around the NGC7332 galaxy group (two bright galaxies) and just one around NGC1156 (Minchin et al. 2010). So surprisingly the galaxy group HI mass function is also not compatible with the global HI function (see also Pisano and Wilcots 2003). From our discussion in section 3.1 we suggest that a significant fraction of the galaxies that contribute to the low mass end of the HI mass function are rather isolated and have nothing to do with brighter galaxies. These isolated low mass galaxies have "appeared" in our global HI mass function, but they are not found in HI mass functions constructed exclusively from group and cluster galaxies. Much larger samples of HI detected galaxies are required to investigate the clustering properties of low mass galaxies compared to higher mass galaxies.

The complete AGES source catalogue is publicly available at the following link http://www.naic.edu/$\sim$ages/public\_data.html.

\section{Optical identifications in the NGC7448 volume}
Identifying the optical counter part to an HI source is not straight forward given the 3.5 arc min beam size and the fact that galaxies cluster. Often there is ambiguity about the optical source and the possibility that more than one source is contributing to the detected HI signal. We have used NED to search for optical counterparts to our HI detections. Most important, though not definitive, is both a close spatial association and a close match in velocity. 51 objects in NED have optical velocities within 100 km s$^{-1}$ of the HI velocity and a spatial separation from the HI position of less than 3 arc min. The median difference between the optical and HI position for these galaxies is $0.3(\pm0.1)$ arc min. This is comparable with values of 0.4 and 0.3 arc min for previous AGES data cubes centred on NGC1156 and NGC7332 respectively (Minchin et al. 2010). The mean difference between the optical and HI Hubble velocities for these 51 galaxies is $12(\pm4)$ km s$^{-1}$, so we assume that these are secure optical identifications. A further 26 galaxies have only one optical possibility from NED. For these galaxies the median difference between the optical and HI position is $0.7(\pm0.7)$ arc min, larger than the separation for the velocity confirmed sample and we cannot be so confident that we have the correct optical source. All the other HI detections have numerous possible optical counterparts and without better resolution we are not able to make a definite association. 

One problem with this particular AGES field is that the optical coverage of the area is not uniform, SDSS only covers part of the field. Where there is SDSS data there is often a number of choices for an optical counterpart. Where there is no SDSS data there is often nothing listed in NED, but a clear candidate is seen when the DSS image is inspected. There are just 5 objects that have no obvious optical counterpart after checking NED and DSS, but all have velocities greater than 13,000 km/s and are relatively low signal-to-noise ratio detections. Any one of the faint 'blobs' in these fields could be the HI source. With the possible exception of AGES J230843+161919 we find no HI detections that have no candidate optical analogue (see also Pisano et al. 2007). It appears that atomic hydrogen is extremely efficient at turning itself into stars.

\section{Conclusions}
Deep blind HI surveys of the sky can reveal galaxies that have gone undetected in other surveys with the added advantage of providing a redshift measurement. In this paper we have described how multi-beam data can be used to both investigate in detail a nearby galaxy group and use galaxies in the surveyed volume to make a measurement of the HI mass function and cosmic HI density. Our main results are:
\begin{enumerate}
\item A large fraction of galaxy group atomic hydrogen may lie between and not within the galaxies.
\item Too few HI-rich dwarf galaxies are detected to be consistent with current models of galaxy formation.
\item  If they had observed this area of sky other wide area blind HI surveys, HIPASS and ALFALFA, would have detected only 5\% and 43\% respectively of the galaxies we detect, missing a large fraction of the atomic gas in this volume. 
\item The global HI mass function is different to that of groups and evolved clusters;
\item We measure a comparatively steep low mass slope to the global HI mass function of $\alpha=-1.53$;
\item Our measured cosmic density of atomic hydrogen is $\Omega_{HI}=5.3 \times 10^{-4}$ h$^{-1}_{72}$
\item We infer a cosmic HI density a factor of 1.7 times higher than that found by HIPASS.
\end{enumerate}

\vspace{0.5cm}
\noindent
{\bf ACKNOWLEDGEMENTS} \\

The Arecibo Observatory is part of the National Astronomy and Ionosphere Center, which is operated by Cornell University under a cooperative agreement with the National Science Foundation. 

This research has made use of the NASA/IPAC Extragalactic Database (NED) which is operated by the Jet Propulsion Laboratory, California Institute of Technology, under contract with the National Aeronautics and Space Administration. 

This research has made use of data obtained from the
SuperCOSMOS Science Archive, prepared and hosted by
the Wide Field Astronomy Unit, Institute for Astronomy,
University of Edinburgh, which is funded by the UK Sci-
ence and Technology Facilities Council. 

The Digitized Sky Surveys were produced at the Space Telescope Science Institute under U.S. Government grant NAG W-2166. The images of these surveys are based on photographic data obtained using the Oschin Schmidt Telescope on Palomar Mountain and the UK Schmidt Telescope. The plates were processed into the present compressed digital form with the permission of these institutions.
The Second Palomar Observatory Sky Survey (POSS-II) was made by the California Institute of Technology with funds from the National Science Foundation, the National Geographic Society, the Sloan Foundation, the Samuel Oschin Foundation, and the Eastman Kodak Corporation.
The Oschin Schmidt Telescope is operated by the California Institute of Technology and Palomar Observatory.
The UK Schmidt Telescope was operated by the Royal Observatory Edinburgh, with funding from the UK Science and Engineering Research Council (later the UK Particle Physics and Astronomy Research Council), until 1988 June, and thereafter by the Anglo-Australian Observatory. The blue plates of the southern Sky Atlas and its Equatorial Extension (together known as the SERC-J), as well as the Equatorial Red (ER), and the Second Epoch [red] Survey (SES) were all taken with the UK Schmidt.

Funding for the SDSS and SDSS-II has been provided by the Alfred P. Sloan Foundation, the Participating Institutions, the National Science Foundation, the U.S. Department of Energy, the National Aeronautics and Space Administration, the Japanese Monbukagakusho, the Max Planck Society, and the Higher Education Funding Council for England. The SDSS Web Site is http://www.sdss.org/.
The SDSS is managed by the Astrophysical Research Consortium for the Participating Institutions. The Participating Institutions are the American Museum of Natural History, Astrophysical Institute Potsdam, University of Basel, University of Cambridge, Case Western Reserve University, University of Chicago, Drexel University, Fermilab, the Institute for Advanced Study, the Japan Participation Group, Johns Hopkins University, the Joint Institute for Nuclear Astrophysics, the Kavli Institute for Particle Astrophysics and Cosmology, the Korean Scientist Group, the Chinese Academy of Sciences (LAMOST), Los Alamos National Laboratory, the Max-Planck-Institute for Astronomy (MPIA), the Max-Planck-Institute for Astrophysics (MPA), New Mexico State University, Ohio State University, University of Pittsburgh, University of Portsmouth, Princeton University, the United States Naval Observatory, and the University of Washington.

\vspace{0.5cm}
\noindent
{\bf REFERENCES} \\
Auld R. et al., 2006, MNRAS, 371, 1617 (Paper I) \\
Barnes D et al., 2001, MNRAS, 322, 486 \\
Bell E., McIntosh D., Katz N. and Weinberg M., 2003, ApJL, 585, L117 \\
Chamaraux P., Balkowski C., and Gerard E., 1980, A\&A, 83, 38  \\
Chynoweth K., Langston G., Holley-Bockelmann K. and Lockman F., AJ, 138, 287 \\
Cole S. et al., 2001, MNRAS, 326, 255  \\
Cortese, L., Gavazzi G., Iglesias-Paramo J. and Carrasco, L. 2004, A\&A, 425, 429  \\
Cortese L. et al., 2008, MNRAS, 383, 1519 (Paper II) \\
Davies J. et al., 2010, A\&A, 518, 48 \\
Davies J. et al., 2004, MNRAS, 349, 922 \\
de Blok E., Zwaan M., Dijkstra M., Briggs F. and Freeman K., 2002, A\&A, 382, 43 \\
Doyle M. et al., 2005, MNRAS, 361, 34 \\
Duprie K. and Schneider S., 1996, AJ, 112, 937 \\
Felten J., 1977, AJ, 82, 861  \\
Giovanelli R. et al., 2005, AJ, 130, 2598 \\
Giovanelli R. et al., 2007, AJ, 133, 2569 \\
Gunn J.E., Gott J.R., 1972, ApJ, 176, 1  \\
Hambly N., Read M., Holliman M., Cross N., Collins R. and Mann R., 2009, ASPC, 411, 381 \\
Haynes M., 1981, AJ, 86, 1126 \\
Haynes M., Giovanelli R. and Chincarini G., 1984, ARAA, 22, 445 \\
Haynes M. P., and Giovanelli R., 1986, ApJ, 306, 466 \\
Kennicutt R., Roettiger K., Keel W., van der Hulst J. and Hummel E., 1987, AJ, 93, 1011 \\
Klypin A., Kravtsov A., Valenzuela O. and Prada F., 1999, ApJ, 522, 82 \\ 
Kovac K., Oosterloo T. and van der Hulst J., 2009, MNRAS, 400, 743 \\
Li J. and Seaquist E., 1994, AJ, 107, 1953 \\
Martin A., Papastergis E., Giovanelli R., Haynes M., Springob, C. and Stierwalt S., 2010, ApJ, submitted (arXiv:1008.5107) \\
Meyer M. et al., 2004, MNRAS, 350, 1195 \\
Minchin R. et al., 2006, in American Astronomical Society Meeting Abstracts, 95.03 \\ 
Minchin R. et al., 2010, ApJ, 140, 1093 (Paper III) \\
Moore, B., Katz, N., Lake, G., Dressler, A., and Oemler, A. 
1996, Nature, 379, 613 \\
Mulchaey J., 2000, ARA\&A, 38, 289 \\
Mulchaey J., Davis D., Mushotzky R. and Burstein D., 2003, ApJS, 145, 39 \\
Pisano D., Barnes D., Gibson B., Stavely-Smith L., Freeman K. and Kilborn V., 2007, ApJ, 662, 959 \\
Pisano D. and Wilcots E., 2003, ApJ, 584, 228 \\
Pisano D., Wakker B., Wilcots E. and Fabian D., 2004, AJ, 127, 199 \\
Putman et al. 2002, AJ, 123, 873 \\
Rosenberg J. and Schneider S., 2002, ApJ, 567, 247  \\
Saintonge A., 2007, AJ, 133, 2087 \\
Salpeter E. and Hoffman G., 1996, ApJ, 465, 595 \\
Schlegel D., Finkbeiner D. and Davis M., 1998, ApJ, 500, 525 \\
Solanes J. M. et al., 2001, ApJ, 548, 97 \\
Spergel D. et al., 2007, ApJS, 170, 377 \\
Swaters R., van Albada T., van der Hulst J. and Sancisi R., 2002, A\&A, 390, 829 \\
Takase B., 1980, PASJ, 32, 605 \\
Taylor R., 2010, PhD thesis, Cardiff University \\
Thomas H., Dunne L., Clemens M., Alexander P, Eales S, Green D. and James A., 2002, MNRAS, 331, 853 \\
Toomre A. and Toomre J., 1972, ApJ, 178, 623 \\
Trentham N., Tully B. and Verheijen M., 2001, MNRAS, 325, 385 \\
Tully R., 1987, ApJ, 321, 280 \\
van Driel W. et al., 1992, A\&A, 259, 71 \\
Verheijen M., Trentham N., Tully R. and Zwaan, M. A., 2000, In	
	'Mapping the Hidden Universe: The Universe behind the Milky Way - The Universe in HI', ASP Conference Proceedings, Vol. 218, edited by R. C. Kraan-Korteweg, P. A. Henning, and H. Andernach. Astronomical Society of the Pacific, ISBN 1-58381-050-1, p.263 \\
Vollmer B., Cayatte V., Balkowski C., and Duschl W. J.
2001, ApJ, 561, 708 \\
Yun M., Ho P. and Lo K., ApJ, 411, L17 \\
Zwaan M., 2001, MNRAS, 325, 1142 \\
Zwaan M, Meyer M., Staveley-Smith L. and Webster R., 2005, MNRAS, 359, L30 \\

\end{document}